\documentclass[12pt,a4paper]{article}
\usepackage{ifthen} % for conditional statements
\newboolean{pdflatex}
\setboolean{pdflatex}{false} % False for eps figures

\newboolean{articletitles}
\setboolean{articletitles}{true} % False removes titles in references

\newboolean{uprightparticles}
\setboolean{uprightparticles}{false} %True for upright particle symbols

\newboolean{inbibliography}
\setboolean{inbibliography}{true} %True once you enter the bibliography

\newboolean{prlstyle}
\setboolean{prlstyle}{false}

\newboolean{wordcount}
\setboolean{wordcount}{false} % False for normal usage; true for wordcount.sh

\usepackage[top=1in, bottom=1.25in, left=1in, right=1in]{geometry}

\usepackage{microtype}
\usepackage{lineno}  % for line numbering during review
\usepackage{xspace} % To avoid problems with missing or double spaces after
\usepackage{caption} %these three command get the figure and table captions automatically small

\usepackage{graphicx}  % to include figures (can also use other packages)
\usepackage{color}
\usepackage{colortbl}
\graphicspath{{./figs/}} % Make Latex search fig subdir for figures

\usepackage{amsmath} % Adds a large collection of math symbols
\usepackage{amssymb}
\usepackage{amsfonts}
\usepackage{upgreek} % Adds in support for greek letters in roman typeset

\newcommand*\patchAmsMathEnvironmentForLineno[1]{%
\expandafter\let\csname old#1\expandafter\endcsname\csname #1\endcsname
\expandafter\let\csname oldend#1\expandafter\endcsname\csname
end#1\endcsname
 \renewenvironment{#1}%
   {\linenomath\csname old#1\endcsname}%
   {\csname oldend#1\endcsname\endlinenomath}%
}
\newcommand*\patchBothAmsMathEnvironmentsForLineno[1]{%
  \patchAmsMathEnvironmentForLineno{#1}%
  \patchAmsMathEnvironmentForLineno{#1*}%
}
\AtBeginDocument{%
\patchBothAmsMathEnvironmentsForLineno{equation}%
\patchBothAmsMathEnvironmentsForLineno{align}%
\patchBothAmsMathEnvironmentsForLineno{flalign}%
\patchBothAmsMathEnvironmentsForLineno{alignat}%
\patchBothAmsMathEnvironmentsForLineno{gather}%
\patchBothAmsMathEnvironmentsForLineno{multline}%
\patchBothAmsMathEnvironmentsForLineno{eqnarray}%
}

\usepackage{hyperref}    % Hyperlinks in references
\usepackage[all]{hypcap} % Internal hyperlinks to floats.

\usepackage{xspace} 
\usepackage{upgreek}

\def\lhcb   {\mbox{LHCb}\xspace}

\def\babar  {\mbox{BaBar}\xspace}
\def\belle  {\mbox{Belle}\xspace}

\def\lhc    {\mbox{LHC}\xspace}

\def\MagUp {\mbox{\em Mag\kern -0.05em Up}\xspace}
\def\MagDown {\mbox{\em MagDown}\xspace}

\ifthenelse{\boolean{uprightparticles}}%
{

 \def\Ppi         {\ensuremath{\uppi}\xspace}                 
                  
 \def\Prho        {\ensuremath{\uprho}\xspace}

 \def\PDelta      {\ensuremath{\Delta}\xspace}                 
 \def\PXi         {\ensuremath{\Xi}\xspace}                 
 \def\PLambda     {\ensuremath{\Lambda}\xspace}                 
 \def\PSigma      {\ensuremath{\Sigma}\xspace}                 
 \def\POmega      {\ensuremath{\Omega}\xspace}                 
 \def\PUpsilon    {\ensuremath{\Upsilon}\xspace}

 \def\PB      {\ensuremath{\mathrm{B}}\xspace}                 
                  
 \def\PD      {\ensuremath{\mathrm{D}}\xspace}

 \def\PK      {\ensuremath{\mathrm{K}}\xspace}

 \def\Pb      {\ensuremath{\mathrm{b}}\xspace}                 
 \def\Pc      {\ensuremath{\mathrm{c}}\xspace}                 
 \def\Pd      {\ensuremath{\mathrm{d}}\xspace}

 \def\Pi      {\ensuremath{\mathrm{i}}\xspace}

 \def\Pp      {\ensuremath{\mathrm{p}}\xspace}

 \def\Ps      {\ensuremath{\mathrm{s}}\xspace}

 \def\thebaroffset{0.0em}
}
{

 \def\Ppi         {\ensuremath{\pi}\xspace}                 
                  
 \def\Prho        {\ensuremath{\rho}\xspace}

 \mathchardef\PDelta="7101
 \mathchardef\PXi="7104
 \mathchardef\PLambda="7103
 \mathchardef\PSigma="7106
 \mathchardef\POmega="710A
 \mathchardef\PUpsilon="7107
                  
 \def\PB      {\ensuremath{B}\xspace}                 
                  
 \def\PD      {\ensuremath{D}\xspace}

 \def\PK      {\ensuremath{K}\xspace}

 \def\Pb      {\ensuremath{b}\xspace}                 
 \def\Pc      {\ensuremath{c}\xspace}                 
 \def\Pd      {\ensuremath{d}\xspace}

 \def\Pi      {\ensuremath{i}\xspace}

 \def\Pp      {\ensuremath{p}\xspace}

 \def\Ps      {\ensuremath{s}\xspace}

 \def\thebaroffset{0.18em}
}
\newcommand{\offsetoverline}[2][\thebaroffset]{\kern #1\overline{\kern -#1 #2}}%

\makeatletter
\ifcase \@ptsize \relax% 10pt
  \newcommand{\miniscule}{\@setfontsize\miniscule{4}{5}}% \tiny: 5/6
\or% 11pt
  \newcommand{\miniscule}{\@setfontsize\miniscule{5}{6}}% \tiny: 6/7
\or% 12pt
  \newcommand{\miniscule}{\@setfontsize\miniscule{5}{6}}% \tiny: 6/7
\fi
\makeatother

\DeclareRobustCommand{\optbar}[1]{\shortstack{{\miniscule (\rule[.5ex]{1.25em}{.18mm})}
  \\ [-.7ex] $#1$}}

\def\dquark    {{\ensuremath{\Pd}}\xspace}
\def\dquarkbar {{\ensuremath{\overline \dquark}}\xspace}

\def\squark    {{\ensuremath{\Ps}}\xspace}

\def\cquark    {{\ensuremath{\Pc}}\xspace}

\def\bquark    {{\ensuremath{\Pb}}\xspace}
\def\bquarkbar {{\ensuremath{\overline \bquark}}\xspace}

\def\pion   {{\ensuremath{\Ppi}}\xspace}
\def\piz    {{\ensuremath{\pion^0}}\xspace}
\def\pip    {{\ensuremath{\pion^+}}\xspace}
\def\pim    {{\ensuremath{\pion^-}}\xspace}

\def\rhomeson {{\ensuremath{\Prho}}\xspace}
\def\rhoz     {{\ensuremath{\rhomeson^0}}\xspace}

\def\kaon    {{\ensuremath{\PK}}\xspace}
\def\Kbar    {{\ensuremath{\offsetoverline{\PK}}}\xspace}

\def\KorKbar {\kern \thebaroffset\optbar{\kern -\thebaroffset \PK}{}\xspace}

\def\Kp      {{\ensuremath{\kaon^+}}\xspace}
\def\Km      {{\ensuremath{\kaon^-}}\xspace}
\def\Kpm     {{\ensuremath{\kaon^\pm}}\xspace}
\def\Kmp     {{\ensuremath{\kaon^\mp}}\xspace}

\def\Kstarz  {{\ensuremath{\kaon^{*0}}}\xspace}
\def\Kstarzb {{\ensuremath{\Kbar{}^{*0}}}\xspace}
\def\Kstar   {{\ensuremath{\kaon^*}}\xspace}

\newcommand{\khigh}{\ensuremath{K^*_0(1430)^0}\xspace}

\def\D       {{\ensuremath{\PD}}\xspace}

\def\DorDbar {\kern \thebaroffset\optbar{\kern -\thebaroffset \PD}\xspace}
\def\Dz      {{\ensuremath{\D^0}}\xspace}

\def\Dm      {{\ensuremath{\D^-}}\xspace}

\def\Dmp     {{\ensuremath{\D^\mp}}\xspace}

\def\Dstarp  {{\ensuremath{\D^{*+}}}\xspace}

\def\Dsm     {{\ensuremath{\D^-_\squark}}\xspace}
\def\Dspm    {{\ensuremath{\D^{\pm}_\squark}}\xspace}
\def\Dsmp    {{\ensuremath{\D^{\mp}_\squark}}\xspace}

\def\B       {{\ensuremath{\PB}}\xspace}
\def\Bbar    {{\ensuremath{\offsetoverline{\PB}}}\xspace}

\def\BorBbar {\kern \thebaroffset\optbar{\kern -\thebaroffset \PB}\xspace}
\def\Bz      {{\ensuremath{\B^0}}\xspace}

\def\Bd      {{\ensuremath{\B^0}}\xspace}
\def\Bs      {{\ensuremath{\B^0_\squark}}\xspace}

\def\Bdb     {{\ensuremath{\Bbar{}^0}}\xspace}

\def\Bq      {{\ensuremath{\B^0_{(\squark)}}}\xspace}
\def\Bqb     {{\ensuremath{\Bbar{}^0_{(\squark)}}}\xspace}

\def\Y#1S{\ensuremath{\PUpsilon{(#1S)}}\xspace}

\def\proton      {{\ensuremath{\Pp}}\xspace}

\def\Lz          {{\ensuremath{\PLambda}}\xspace}

\def\LorLbar     {\kern \thebaroffset\optbar{\kern -\thebaroffset \PLambda}\xspace}

\def\Lc          {{\ensuremath{\Lz^+_\cquark}}\xspace}

\def\Lb           {{\ensuremath{\Lz^0_\bquark}}\xspace}

\def\BF         {{\ensuremath{\mathcal{B}}}\xspace}
\def\BR         {\BF}

\newcommand{\decay}[2]{\mbox{\ensuremath{#1\!\to #2}}\xspace}         % {\Pa}{\Pb \Pc}

\def\to                 {\ensuremath{\rightarrow}\xspace}

\def\CP                {{\ensuremath{C\!P}}\xspace}

\newcommand{\GL}{{\ensuremath{\Gamma_{\mathrm{ L}}}}\xspace}
\newcommand{\GH}{{\ensuremath{\Gamma_{\mathrm{ H}}}}\xspace}

\newcommand{\Kpi}{\ensuremath{(K\pi)}\xspace}
\newcommand{\Kpia}{\ensuremath{(\Kp\pim)}\xspace}
\newcommand{\Kpib}{\ensuremath{(\Km\pip)}\xspace}
\newcommand{\KpiKpi}{\ensuremath{(\Kp\pim\Km\pip)}\xspace}

\def\BsKstKst    {\decay{\Bs}{\Kstarz\Kstarzb}}
\def\BdKstKst    {\decay{\Bd}{\Kstarz\Kstarzb}}
\def\BqKstKst    {\decay{\Bq}{\Kstarz\Kstarzb}}
\def\BsKpiKpi    {\decay{\Bs}{(\Kp\pim)(\Km\pip)}}
\def\BdKpiKpi    {\decay{\Bd}{(\Kp\pim)(\Km\pip)}}
\def\BqKpiKpi    {\decay{\Bq}{(\Kp\pim)(\Km\pip)}}

\def\AT#1     {\ensuremath{A_{\mathrm{T}}^{#1}}\xspace}           % 2

\def\C#1      {\ensuremath{\mathcal{C}_{#1}}\xspace}                       % 9
\def\Cp#1     {\ensuremath{\mathcal{C}_{#1}^{'}}\xspace}                    % 7
\def\Ceff#1   {\ensuremath{\mathcal{C}_{#1}^{\mathrm{(eff)}}}\xspace}        % 9  
\def\Cpeff#1  {\ensuremath{\mathcal{C}_{#1}^{'\mathrm{(eff)}}}\xspace}       % 7
\def\Ope#1    {\ensuremath{\mathcal{O}_{#1}}\xspace}                       % 2
\def\Opep#1   {\ensuremath{\mathcal{O}_{#1}^{'}}\xspace}                    % 7

\newcommand{\nospaceunit}[1]{\ensuremath{\text{#1}}}       
\newcommand{\aunit}[1]{\ensuremath{\text{\,#1}}}       
\newcommand{\tev}{\aunit{Te\kern -0.1em V}\xspace}
\newcommand{\gev}{\aunit{Ge\kern -0.1em V}\xspace}
\newcommand{\mev}{\aunit{Me\kern -0.1em V}\xspace}
\newcommand{\kev}{\aunit{ke\kern -0.1em V}\xspace}
\newcommand{\ev}{\aunit{e\kern -0.1em V}\xspace}
\newcommand{\mevc}{\ensuremath{\aunit{Me\kern -0.1em V\!/}c}\xspace}
\newcommand{\gevc}{\ensuremath{\aunit{Ge\kern -0.1em V\!/}c}\xspace}
\newcommand{\mevcc}{\ensuremath{\aunit{Me\kern -0.1em V\!/}c^2}\xspace}
\newcommand{\gevcc}{\ensuremath{\aunit{Ge\kern -0.1em V\!/}c^2}\xspace}
\def\mm   {\aunit{mm}\xspace}

\def\mum  {\ensuremath{\,\upmu\nospaceunit{m}}\xspace}

\def\fb   {\ensuremath{\aunit{fb}}\xspace}
\def\invfb   {\ensuremath{\fb^{-1}}\xspace}

\def\ps   {\ensuremath{\aunit{ps}}\xspace}

\def\invps{\ensuremath{\ps^{-1}}\xspace}

\newcommand{\stat}{\aunit{(stat)}\xspace}
\newcommand{\syst}{\aunit{(syst)}\xspace}

\newcommand{\chisq}{\ensuremath{\chi^2}\xspace}
\newcommand{\chisqndf}{\ensuremath{\chi^2/\mathrm{ndf}}\xspace}
\newcommand{\chisqip}{\ensuremath{\chi^2_{\text{IP}}}\xspace}

\def\deriv {\ensuremath{\mathrm{d}}}

\def\gsim{{~\raise.15em\hbox{$>$}\kern-.85em
          \lower.35em\hbox{$\sim$}~}\xspace}
\def\lsim{{~\raise.15em\hbox{$<$}\kern-.85em
          \lower.35em\hbox{$\sim$}~}\xspace}

\newcommand{\Real}{\ensuremath{\mathcal{R}e}\xspace}

\def\PDF {PDF\xspace}

\def\sPlot{\mbox{\em sPlot}\xspace}

\def\pt         {\ensuremath{p_{\mathrm{T}}}\xspace}

\def\ptot       {\ensuremath{p}\xspace}

\def\dllkpi     {\ensuremath{\mathrm{DLL}_{\kaon\pion}}\xspace}

\def\dllpk      {\ensuremath{\mathrm{DLL}_{\proton\kaon}}\xspace}

\def\evtgen     {\mbox{\textsc{EvtGen}}\xspace}

\def\geant      {\mbox{\textsc{Geant4}}\xspace}

\def\photos     {\mbox{\textsc{Photos}}\xspace}

\def\pythia     {\mbox{\textsc{Pythia}}\xspace}

\xspace

\def\tell1  {TELL1\xspace}
\def\ukl1   {UKL1\xspace}

\newcommand{\swave}{\mbox{{S--wave}}\xspace}
\newcommand{\pwave}{\mbox{{P--wave}}\xspace}
\newcommand{\dwave}{{D--wave}\xspace}

\newcommand{\figref}[1]{Fig.~\ref{#1}}
\newcommand{\tabref}[1]{Table~\ref{#1}}

\newcommand{\secref}[1]{Sect.~\ref{#1}}
\renewcommand{\eqref}[1]{Eq.~(\ref{#1})}
\usepackage{cite} % Allows for ranges in citations
\usepackage{mciteplus}

\usepackage{longtable} % only for template; not usually to be used in PAPERs
\usepackage{rotating}
\usepackage{arydshln} % for dashed lines in tables
\usepackage{multirow}
\usepackage{diagbox}
\usepackage{booktabs}

\usepackage{amsmath}
\usepackage[makeroom]{cancel}

\usepackage{mciteplus}

\usepackage{lineno}
\def\papertitle{Amplitude analysis of the $\Bq \to \Kstarz \Kstarzb$ decays and measurement of the branching fraction of the $\Bd \to \Kstarz \Kstarzb$ decay} % Latex formatted title
\def\papercopyright{\the\year\ CERN for the benefit of the LHCb collaboration} % new since 9/Apr/2018
\def\paperlicence{CC-BY-4.0 licence}
\def\paperlicenceurl{https://creativecommons.org/licenses/by/4.0/}

\begin{document}
%%%%%%%%%%%%%%%%%%%%%%%%%
%%%%% Title     %%%%%%%%%
%%%%%%%%%%%%%%%%%%%%%%%%%
\renewcommand{\thefootnote}{\fnsymbol{footnote}}
\setcounter{footnote}{1}

% %%%%%%% CHOOSE TITLE PAGE--------
% $Id: title-LHCb-PAPER.tex 78711 2015-08-06 07:54:32Z apuignav $
% ===============================================================================
% Purpose: LHCb-PAPER journal paper title page template
% Author:
% Created on: 2010-09-25
% ===============================================================================

%%%%%%%%%%%%%%%%%%%%%%%%%
%%%%%  TITLE PAGE  %%%%%%
%%%%%%%%%%%%%%%%%%%%%%%%%
\begin{titlepage}
\pagenumbering{roman}

% Header ---------------------------------------------------
\vspace*{-1.5cm}
\centerline{\large EUROPEAN ORGANIZATION FOR NUCLEAR RESEARCH (CERN)}
\vspace*{1.5cm}
\noindent
\begin{tabular*}{\linewidth}{lc@{\extracolsep{\fill}}r@{\extracolsep{0pt}}}
\ifthenelse{\boolean{pdflatex}}% Logo format choice
{\vspace*{-2.7cm}\mbox{\!\!\!\includegraphics[width=.14\textwidth]{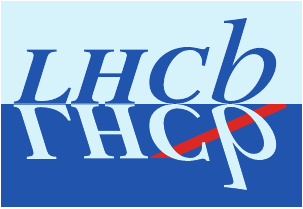}} & &}%
{\vspace*{-1.2cm}\mbox{\!\!\!\includegraphics[width=.12\textwidth]{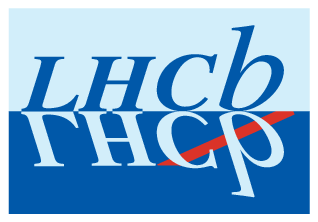}} & &}%
\\
 & & CERN-EP-2019-063 \\  % ID
 & & LHCb-PAPER-2019-004 \\  % ID
%& & \today \\ % Date - Can also hardwire e.g.: 23 March 2010
  & & July 16, 2019 \\
%  & & May 16, 2019 \\
% & & version 6.1  \\
% not in paper \hline
\end{tabular*}

\vspace*{2.0cm}

% Title --------------------------------------------------
{\normalfont\bfseries\boldmath\huge
\begin{center}
  \papertitle 
\end{center}
}

\vspace*{1.0cm}

% Authors -------------------------------------------------
\begin{center}
The LHCb collaboration\footnote{Authors are listed at the end of this letter.}
\end{center}

\vspace{\fill}

% Abstract -----------------------------------------------
\begin{abstract}
\noindent
The $B^0 \to \Kstarz \Kstarzb$ and $B^0_s \to \Kstarz \Kstarzb$ decays are studied using proton-proton collision data corresponding to an integrated luminosity of 3\invfb. An untagged and time-integrated amplitude analysis of $\Bq \to \Kpia \Kpib$ decays in two-body invariant mass regions of 150\mevcc around the \Kstarz mass is performed. A stronger longitudinal polarisation fraction in the ${B^0\to \Kstarz \Kstarzb}$ decay, ${f_L = 0.724 \pm 0.051 \,({\rm stat}) \pm 0.016 \,({\rm syst})}$, is observed as compared to ${f_L = 0.240 \pm 0.031 \,({\rm stat}) \pm 0.025 \,({\rm syst})}$ in the ${B^0_s\to \Kstarz \Kstarzb}$ decay. The ratio of branching fractions of the two decays is measured and used to determine $\BF (\BdKstKst) =  (8.0 \pm 0.9 \,({\rm stat}) \pm 0.4 \,({\rm syst})) \times 10^{-7}$.
\end{abstract}

\vspace*{2.0cm}

\begin{center}
 Published in JHEP 07 (2019) 032
\end{center}

\vspace{\fill}

{\footnotesize
\centerline{\copyright~\papercopyright. \href{\paperlicenceurl}{\paperlicence}.}}
\vspace*{2mm}

\end{titlepage}

%%%%%%%%%%%%%%%%%%%%%%%%%%%%%%%%
%%%%%  EOD OF TITLE PAGE  %%%%%%
%%%%%%%%%%%%%%%%%%%%%%%%%%%%%%%%

%  empty page follows the title page ----
\newpage
\setcounter{page}{2}
\mbox{~}

\cleardoublepage

% %%%%%%%%%%%%% ---------

\renewcommand{\thefootnote}{\arabic{footnote}}
\setcounter{footnote}{0}

%\nocite{*}   irrelevant

%%%%%%%%%%%%%%%%%%%%%%%%%
%%%%% Main text %%%%%%%%%
%%%%%%%%%%%%%%%%%%%%%%%%%

\pagestyle{plain} % restore page numbers for the main text
\setcounter{page}{1}
\pagenumbering{arabic}

\section{Introduction}
\label{sec:introduction}

The \BdKstKst decay is a Flavour-Changing Neutral Current (FCNC) process.\footnote{Throughout the text charge conjugation is implied, \Kpi indicates either a \Kpia or a \Kpib pair, \Bq indicates either a \Bd or a \Bs meson and \Kstarz refers to the $\Kstar (892)^0$ resonance, unless otherwise stated.} In the Standard Model (SM) this type of processes is forbidden at tree level and occurs at first order through loop penguin diagrams. Hence, FCNC processes are considered to be excellent probes for physics beyond the SM, since contributions mediated by heavy particles, contemplated in these theories, may produce effects measurable with the current sensitivity.

Evidence of the \BdKstKst decay has been found by the \babar collaboration~\cite{babar} with a measured yield of  ${33.5}^{+9.1}_{-8.1}$ decays. An untagged time-integrated analysis was presented finding a branching fraction of $\BF=(1.28^{+0.35}_{-0.30} \pm 0.11)\times {10}^{-6} $ and a longitudinal polarisation fraction of $f_L = {0.80}^{+0.11}_{-0.12} \pm 0.06$. In untagged time-integrated analyses the distributions for \Bd and \Bdb decays are assumed to be identical and summed, so that they can be fitted with a single amplitude. However, if \CP-violation effects are present, the distribution is given by the incoherent sum of the two contributions. The \belle collaboration also searched for this decay~\cite{Chiang:2010ga} and a branching fraction of $\BF=(0.26^{+0.33+0.10}_{-0.29-0.07}) \times 10^{-6}$ was measured, disregarding \swave contributions. There is a $2.2$ standard-deviations difference between the branching fraction measured by the two experiments. The predictions of factorised QCD (QCDF) are $\BF= (0.6^{+0.1+0.5}_{-0.1-0.3})\times 10^{-6}$ and $f_L=0.69^{+0.01+0.34}_{-0.01-0.27}$~\cite{beneke}. Perturbative QCD predicts $\BF=(0.64^{+0.24}_{-0.23})\times 10^{-6}$~\cite{LHCb-PAPER-2011-012}.\footnote{This reference considers two scenarios for its predictions, both giving compatible results. Only the first scenario considered therein is quoted here.} These theoretical predictions agree with the experimental results within the large uncertainties. The measurement of $f_L$ agrees with the na\"ive hypothesis, based on the quark helicity conservation and the $V$$-$$A$ nature of the weak interaction, that charmless decays into pairs of vector mesons ($VV$) should be strongly longitudinally polarised. See, for example, the {\em Polarization in B Decays} review in Ref.~\cite{pdg-2018}.

\begin{figure}
\begin{center}
\includegraphics[width=0.4\textwidth]{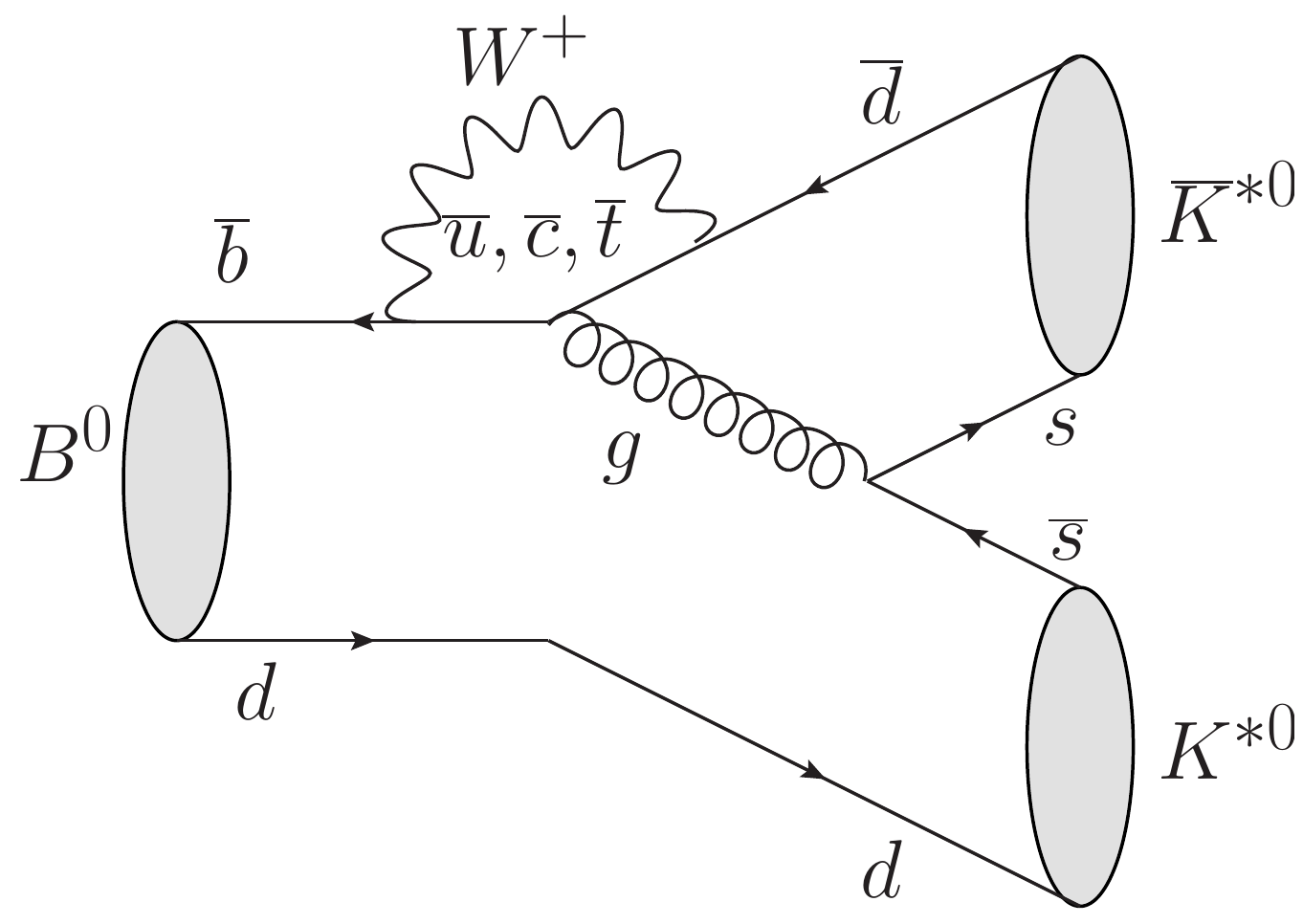}
\includegraphics[width=0.4\textwidth]{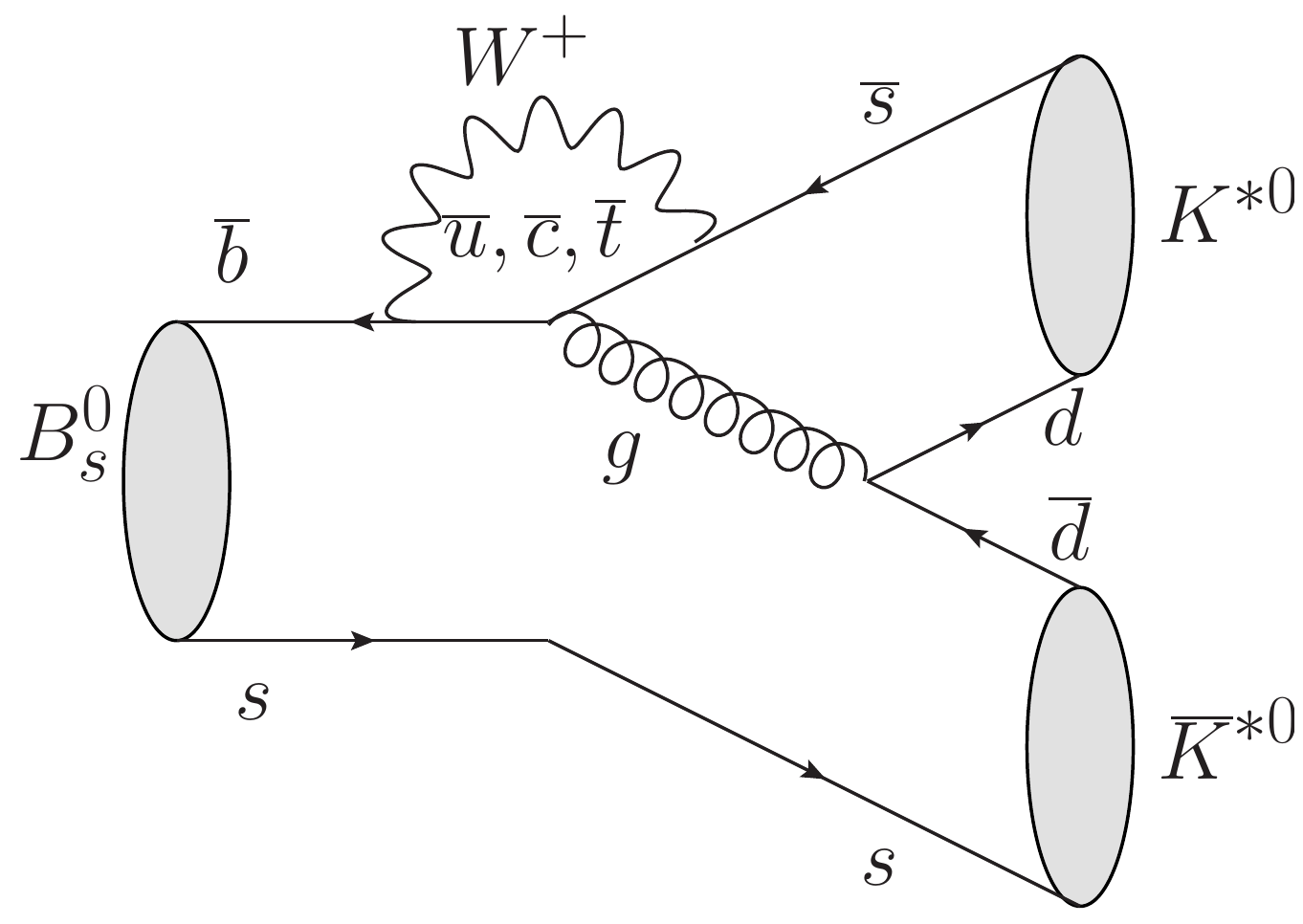}
\caption{Leading order Feynman diagrams for the \BdKstKst and \BsKstKst decays. Both modes are dominated by a gluonic-penguin diagram.\label{fig:feynman-diagrams}}
\end{center}
\end{figure}

The \BsKstKst decay was first observed by the \lhcb experiment with early \lhc data~\cite{LHCb-PAPER-2011-012}. A later untagged time-integrated study, with data corresponding to $1\invfb$ of integrated luminosity, measured ${\BF = (10.8 \pm  2.1 \pm 1.5)\times {10}^{-6}}$ and ${f_L= 0.201 \pm 0.057 \pm 0.040}$~\cite{LHCb-PAPER-2014-068}. More recently, a complete \CP-sensitive time-dependent analysis of \BsKpiKpi decays in the \Kpi mass range from 750 to $1600\mevcc$ has been published by \lhcb~\cite{LHCb-PAPER-2017-048}, with data corresponding to $3\invfb$ of integrated luminosity. A determination of ${f_L = 0.208 \pm 0.032 \pm 0.046}$ was performed as well as the first measurements of the mixing-induced \CP-violating phase $\phi_s^{\dquark\dquarkbar}$ and of the direct \CP asymmetry parameter $|\lambda|$. These \lhcb analyses of \BsKpiKpi decays lead to three conclusions: firstly, within their uncertainties, the measured observables are compatible with the absence of \CP violation; secondly, a low polarisation fraction is found; finally, a large \swave contribution, as much as $60\%$, is measured in the $150\mevcc$ window around the \Kstarz mass. The low longitudinal polarisation fraction shows a tension with the prediction of QCDF ($f_L=0.63^{+0.42}_{-0.29}$~\cite{beneke}) and disfavours the hypothesis of strongly longitudinally polarised $VV$ decays. Theoretical studies try to explain the small longitudinal polarisation with mechanisms such as contributions from annihilation processes~\cite{Kagan:2004uw,beneke}. It is intriguing that the two channels \BdKstKst and \BsKstKst, which are related by U-spin symmetry, implying the exchange of \dquark and \squark quarks as displayed in~\figref{fig:feynman-diagrams}, show such different polarisations. A comprehensive theory review on polarisation of charmless $VV$ neutral \B-meson decays can be found in Ref.~\cite{Kou:2018nap}.

Some authors consider the \BsKstKst decay as a golden channel for a precision test of the CKM phase $\beta_s$~\cite{ciucini}. High-precision analyses of this channel, dominated by the gluonic-penguin diagram, will require to account for subleading amplitudes~\cite{Bediaga:2018lhg,Kou:2018nap}. The study of the \BdKstKst decay allows to control higher-order SM contributions to the \BsKstKst channel employing U-spin symmetry~\cite{ciucini,matias}. In Refs.~\cite{matias,DescotesGenon:2011pb} more precise QCDF predictions, involving the relation between longitudinal branching fractions of the two channels, are made.

In this work, an untagged and time-integrated amplitude analysis of the \BdKpiKpi and \BsKpiKpi decays in  the two-body invariant mass regions of 150\mevcc around the \Kstarz mass is presented, as well as the determination of the \BdKstKst decay branching fraction. The analysis uses data recorded in 2011 and 2012 at centre-of-mass energies of $\sqrt{s}=7$ and $\sqrt{s}=8\tev$, respectively, corresponding to an integrated luminosity of $3\invfb$. 

This paper is organised as follows. In~\secref{sec:amplitude-formalism} the formalism of the decay amplitudes is presented. In~\secref{sec:Detector} a brief description of the \lhcb detector, online selection algorithms and simulation software is given. The selection of \BdKpiKpi and \BsKpiKpi candidates is presented in~\secref{sec:selection}. \secref{sec:4body} describes the maximum-likelihood fit to the four-body invariant-mass spectra and its results. The amplitude analysis and its results are discussed in~\secref{sec:amplitude-analysis}. The estimation of systematic uncertainties is described in~\secref{sec:systematic-uncertainties}, and the determination of the \BdKstKst decay branching fraction relative to the \BsKstKst mode in~\secref{sec:branching}. Finally, the results are summarised and conclusions are drawn in~\secref{sec:conclusion}.

\section{Amplitude analysis formalism}
\label{sec:amplitude-formalism}

\begin{figure}
\begin{center}
\includegraphics[width=0.7\textwidth]{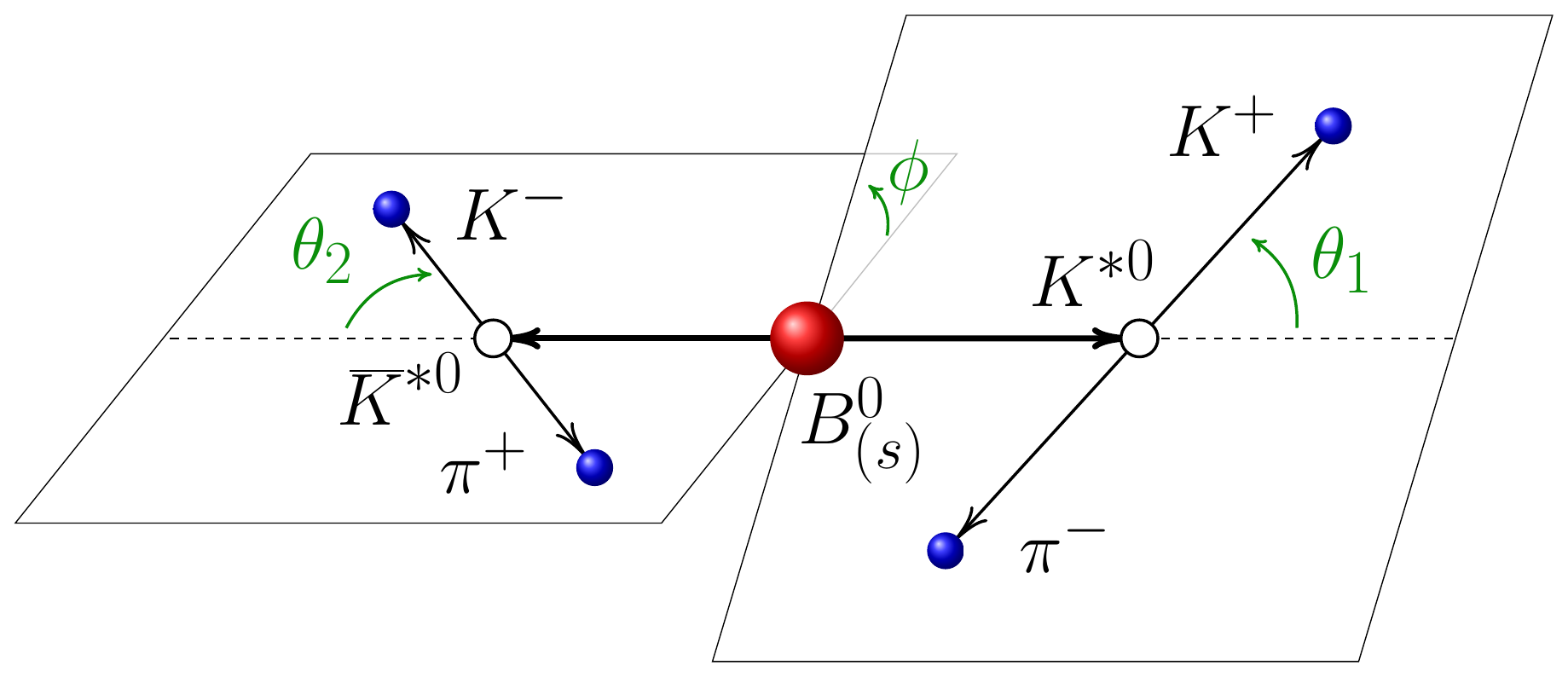}
\caption{Definition of the helicity angles, employed in the angular analysis of the \BqKstKst decays. Each angle is defined in the rest frame of the decaying particle.\label{fig:angles-definition}}
\end{center}
\end{figure}

The \BdKstKst and \BsKstKst modes are weak decays of a pseudoescalar particle into two vector mesons ($\decay{P}{V V}$). The \B-meson decays are followed by subsequent \decay{\Kstarz}{\Kp\pim} and \decay{\Kstarzb}{\Km\pip} decays. The study of the angular distribution employs the helicity angles shown in~\figref{fig:angles-definition}: $\theta_{1(2)}$, defined as the angle between the direction of the $\kaon^{+(-)}$ meson and the direction opposite to the \B-meson momentum in the rest frame of the \Kstarz (\Kstarzb) resonance, and $\phi$, the angle between the decay planes of the two vector mesons in the \B-meson rest frame. From angular momentum conservation, three relative polarisations of the final state are possible for $VV$ final states that correspond to longitudinal ($0$ or $L$), or transverse to the direction of motion and parallel ($\parallel$) or perpendicular ($\perp$) to each other. For the two-body invariant mass of the \Kpia and \Kpib pairs, noted as $m_1\equiv M\Kpia$ and $m_2\equiv M\Kpib$, a range of $150$\mevcc around the known \Kstarz mass~\cite{pdg-2018} is considered. Therefore, \Kpi pairs may not only originate from the spin-1 \Kstarz meson, but also from other spin states. This justifies that, besides the helicity angles, a phenomenological description of the two-body invariant mass spectra, employing the isobar model, is adopted in the analytic model. In the isobar approach, the decay amplitude is modelled as a linear superposition of quasi-two-body amplitudes~\cite{Fleming:1964zz,*Morgan:1968zza,*Herndon:1973yn}.

For the \swave ($J=0$), the \khigh resonance, the possible $K^*_0(700)^0$ (or $\kappa$) and a non-resonant component, $\Kpi_0$, need to be accounted for. This is done using the LASS parameterisation~\cite{LASS}, which is an effective-range elastic scattering amplitude, interfering with the \khigh meson,
\begin{equation}
\label{eq:bwKpiSwave}
 \mathcal{M}_0(m) \propto 
 \frac{m}{q} \left(
 \frac{1}{\cot\delta_{\beta} - i} + e^{2i\delta_{\beta}}\frac{M_0 \Gamma_0(m)}{M_0^2 - m^2 - i M_0 \Gamma_0(m) }
 \right),
\end{equation}
where
\begin{equation}
\label{eq:Gamma0}
 \Gamma_0(m) = \Gamma_0 \frac{M_0}{m} \left ( \frac{q}{q_0} \right )
\end{equation}
\noindent represents the \khigh width. In~\eqref{eq:bwKpiSwave} and~\eqref{eq:Gamma0} $q$ is the \Kpi centre-of-mass decay momentum, and $M_0$, $\Gamma_0$ and $q_0$ are the \khigh mass, width and centre-of-mass decay momentum at the pole, respectively. The effective-range elastic scattering amplitude component depends on
\[
\cot\delta_{\beta} = \frac{1}{a q} + \frac{1}{2}bq,
\]
where $a$ is the scattering length and $b$ the effective range.

For the \pwave ($J=1$), only the $\kaon^*(892)^0$ resonance is considered. Other \pwave resonances, such as $\kaon^*(1410)^0$ or $\kaon^*(1680)^0$, with pole masses much above the fit region, are neglected. Resonances with higher spin, for instance the \dwave $\kaon^*_2(1430)^0$ meson, are negligible in the considered two-body mass range~\cite{LHCb-PAPER-2017-048} and are also disregarded. The \Kstarz amplitude is parameterised with a spin-1 relativistic Breit--Wigner amplitude,
\begin{equation}
  \mathcal{M}_1(m) \propto \frac{m}{q}\frac{M_1 \Gamma_1(m)}{(M_1 ^2 - m^2) - i M_1\Gamma_1(m) }.
\label{eq:propag}
\end{equation}
\noindent The  mass-dependent width is given by
\begin{equation}
  \Gamma_1(m) = \Gamma_1 \frac{M_1}{m} \frac{1+r^2 q_1^2}{1 + r^2 q^2} \left(\frac{q}{q_1}\right)^3,
\end{equation}
\noindent where $M_1$ and $\Gamma_1$ are the \Kstarz mass and width, $r$ is the interaction radius parameterising the centrifugal barrier penetration factor, and $q_1$ corresponds to the centre-of-mass decay momentum at the resonance pole. The values of the mass propagator parameters are summarised in~\tabref{table:inputSwave}.

\begin{table}[t]
\begin{center}
\caption{Parameters of the mass propagators employed in the amplitude analysis.}
\begin{tabular}{@{}c@{}@{}l@{}|c|c}
  & & $(K\pi)_0$  & \Kstarz  \\
  & & $J=0$~\cite{LASS,bdphikst_babar}  & $J=1$~\cite{pdg-2018}  \\
  \hline
   $M_J$ & \phantom{,}[MeV/$c^2$]\phantom{,,}& $1435 \pm \phantom{2}7$ & $895.81 \pm 0.19$ \\
   $\Gamma_{J}$ & \phantom{,}[MeV]  & $\phantom{1}279 \pm 22$ & $\phantom{0}47.4\phantom{1} \pm 0.6\phantom{0}$  \\
   $r$ & \phantom{,}[$c$/GeV]    & $\phantom{00}-$ & $\phantom{00}3.0\phantom{1} \pm 0.5\phantom{0}$ \\
   $a$ & \phantom{,}[$c$/GeV]    & $\phantom{00}1.95 \pm 0.11$ & $\phantom{00}-$ \\
   $b$ & \phantom{,}[$c$/GeV]    & $\phantom{00}1.76 \pm 0.76$ & $\phantom{00}-$  \\
\end{tabular}
\label{table:inputSwave}
\end{center}
\end{table}

The differential decay rate for \Bq mesons\footnote{Charge conjugation is not implied in the rest of this section. For the charge-conjugated mode, \decay{\Bqb}{\Kpia \Kpib}, the decay rate is obtained applying the transformation $A_i \to \eta_i \bar{A}_i$ in~\eqref{eq:allAmplitudes_SV2} where the corresponding \CP eigenvalues, $\eta_i$, are given in~\tabref{tab:amplitudes}.} at production is given by~\cite{london_new, LHCb-PAPER-2014-068},
\begin{align}
\frac{\deriv^5 \Gamma}{\deriv {\rm cos} \, \theta_1 \deriv {\rm cos}\,  \theta_2 \deriv \phi \deriv m_1 \deriv m_2}
&= \frac{9}{8\pi} \Phi_4(m_1,m_2) \left| \sum_{i=1}^{6} A_i g_i(m_1,m_2,\theta_1,\theta_2,\phi) \right|^2 \nonumber \\
&= \sum_{i=1}^{6} \sum_{j\geq i}^6 \Real [A_iA^*_j F_{ij}],
\label{eq:allAmplitudes_SV2}
\end{align}
where $\Phi_4$ is the four-body phase space factor. The index $i$ runs over the first column of~\tabref{tab:amplitudes} where the different decay amplitudes, $A_i \equiv |A_i|e^{i\delta_i}$, and the angular-mass functions, $g_i$, are listed. The angular dependence of these functions is obtained from spherical harmonics as explained in Ref.~\cite{london_new}. For \CP-studies, the \CP-odd, $A_S^+$, and \CP-even, $A_S^-$, eigenstates of the \swave polarisation amplitudes are preferred to the vector-scalar ($VS$) and scalar-vector ($SV$) helicity amplitudes, to which they are related by
\[
A_S^+ = \frac{A_{VS} + A_{SV}}{\sqrt{2}} \quad \text{and} \quad 
A_S^- = \frac{A_{VS} - A_{SV}}{\sqrt{2}}.
\]
The remaining amplitudes, except for $A_\perp$, correspond to \CP-even eigenstates. The contributions can be quantified by the terms $F_{ij}$, defined as
\begin{equation}
\label{eq:Fij}
F_{ij} = \frac{9}{8\pi} \Phi_4 (m_1,m_2)g_i(m_1,m_2,\theta_1,\theta_2,\phi)g_j^*(m_1,m_2,\theta_1,\theta_2,\phi) (2-\delta_{ij}),
\end{equation}
which are normalised according to
\[
\int F_{ij} \deriv m_1 \deriv m_2 \deriv {\rm cos}\,\theta_1 \deriv {\rm cos}\,\theta_1 \deriv \phi = \delta_{ij}.
\]
This condition ensures that $\sum_{i=1}^6 |A_i|^2=1$.

\begin{table}
\caption{Amplitudes, $A_i$, and angle-mass functions, $g_i(m_1,m_2,\theta_1,\theta_2,\phi)$, of the differential decay rate of~\eqref{eq:allAmplitudes_SV2}. In particular, $A_0$, $A_\parallel$ and $A_\perp$ are the longitudinal, parallel and transverse helicity amplitudes of the \pwave whereas $A_S^+$ and $A_S^-$ are the combinations of \CP eigenstate amplitudes of the $SV$ and $VS$ states and $A_{SS}$ is the double \swave amplitude. The table indicates the corresponding \CP eigenvalue, $\eta_i$. The mass propagators, $\mathcal{M}_{0,1}(m)$, are discussed in the text.}
\label{tab:amplitudes}
\begin{center}
\begin{tabular}{c c c c }
\hline
$i$ & $A_i$ & $\phantom{-}\eta_i$ &       $g_i(m_1,m_2,\theta_1,\theta_2,\phi)$  \\
\hline
1 & $A_0$ & $\phantom{-}1$ & $\cos \theta_1 \cos \theta_2 {\cal M}_1 (m_1) {\cal M}_1 (m_2)$ \\
2 & $A_\parallel$ & $\phantom{-}1$ & $\frac{1}{\sqrt{2}} \sin \theta_1 \sin \theta_2 \cos \phi {\cal M}_1 (m_1) {\cal M}_1 (m_2)$ \\
3 & $A_\perp$ & $-1$ & $\frac{i}{\sqrt{2}} \sin \theta_1 \sin \theta_2 \sin \phi {\cal M}_1 (m_1) {\cal M}_1 (m_2)$ \\
4 & $A_S^+$ & $-1$ & $-\frac{1}{\sqrt{6}} (\cos \theta_1 {\cal M}_1 (m_1) {\cal M}_0 (m_2) - \cos \theta_2 {\cal M}_0 (m_1) {\cal M}_1 (m_2))$ \\
5 & $A_S^-$ & $\phantom{-}1$ & $-\frac{1}{\sqrt{6}} (\cos \theta_1 {\cal M}_1 (m_1) {\cal M}_0 (m_2) + \cos \theta_2 {\cal M}_0 (m_1) {\cal M}_1 (m_2))$ \\
6 & $A_{SS}$ & $\phantom{-}1$ & $- \frac{1}{3} {\cal M}_0 (m_1) {\cal M}_0 (m_2)$ \\
\hline
\end{tabular}
\end{center}
\end{table}

The polarisation fractions of the $VV$ amplitudes are defined as
\[
f_{L,\parallel,\perp} = \frac{|A_{0,\parallel, \perp}|^2}{|A_0|^2 + |A_\parallel|^2 + |A_\perp|^2},
\]
where  $A_0$, $A_\parallel$ and $A_\perp$ are the longitudinal, parallel and transverse amplitudes of the \pwave. Therefore, $f_L$ is the fraction of \BqKstKst longitudinally polarised decays. The polarisation fractions are preferred to the amplitude moduli since they are independent of the considered \Kpi mass range. The \pwave amplitudes moduli can always be recovered as
\[
\label{eq:pwavemoduli}
|A_{0,\parallel, \perp}|^2 = (1-|A_S^+|^2-|A_S^-|^2-|A_{SS}|^2)\, f_{L,\parallel,\perp}.
\]

The phase of all propagators is set to be zero at the \Kstarz mass. In addition, a global phase can be factorised without affecting the decay rate setting $\delta_0\equiv 0$. The last two requirements establish the definition of the amplitude phases ($\delta_\parallel$, $\delta_\perp$, $\delta_S^-$, $\delta_S^+$ and $\delta_{SS}$) as the phase relative to that of the longitudinal \pwave amplitude at the \Kstarz mass.

Since \Bq mesons oscillate, the decay rate evolves with time. The time-dependent amplitudes are obtained replacing $A_i \to \mathcal{A}_i (t)$ and $\bar{A}_i \to \bar{\mathcal{A}}_i (t)$ in~\eqref{eq:allAmplitudes_SV2} being
\begin{equation}
\label{eq:amp-time}
\mathcal{A}_i (t) = \left[ g_+ (t) A_i + \eta_i \frac{{\rm q}}{{\rm p}} g_- (t) \bar{A}_i \right] \quad \text{and} \quad
\bar{\mathcal{A}}_i (t) = \left[ \frac{{\rm p}}{{\rm q}} g_- (t) A_i + \eta_i g_+ (t) \bar{A}_i \right], \nonumber
\end{equation}
with
\begin{equation}
g_+ (t) = \frac{1}{2} \left( e^{-\left(i M_{\rm L} + \frac{\GL}{2}\right)t} + e^{-\left(i M_{\rm H} + \frac{\GH}{2}\right)t} \right) \; \text{and} \;
g_- (t) = \frac{1}{2} \left( e^{-\left(i M_{\rm L} + \frac{\GL}{2}\right)t} - e^{-\left(i M_{\rm H} + \frac{\GH}{2}\right)t} \right), \nonumber
\end{equation}
where $\GL$ and $\GH$ are the widths of the light and heavy mass eigenstates of the $\Bq-\Bqb$ system and $M_{\rm L}$ and $M_{\rm H}$ are their masses. The coefficients p and q are the mixing terms that relate the flavour and mass eigenstates,
\begin{equation}
\Bq_{\rm H} = {\rm p} \Bq + {\rm q} \Bqb \quad \text{and} \quad
\Bq_{\rm L} = {\rm p} \Bq - {\rm q} \Bqb. \nonumber
\end{equation}
Masses and widths are often considered in their averages and differences, ${M = (M_{\rm L}+M_{\rm H})/2}$, $\Delta M = M_{\rm L}-M_{\rm H}$, $\Gamma = (\GL+\GH)/2$ and $\Delta \Gamma = \GL-\GH$, in particular in their relation with the mixing phase,
\begin{equation}
\label{eq:mixingphase}
\tan \phi_{(s)}= 2\frac{\Delta M}{\Delta \Gamma} \left(1-\frac{|{\rm q}|}{|{\rm p}|}\right).
\end{equation}

In this analysis, no attempt is made to identify the flavour of the initial \Bq meson and time-integrated spectra are considered. Consequently, the selected candidates correspond to untagged and time-integrated decay rates and there is no sensitivity to direct and mixing-induced \CP violation. Moreover, since the origin of phases is set in a \CP-even eigenstate ($\delta_0=0$), for the \CP-odd eigenstates, the untagged time-integrated decay is only sensitive to the phase difference $\delta_\perp - \delta_S^+$. The present experimental knowledge is compatible with small \CP violation in mixing~\cite{HFLAV} and with the absence of direct \CP violation in the \BsKpiKpi system~\cite{LHCb-PAPER-2017-048}.

The dependence of the decay rate in an untagged and time-integrated analysis of a \Bq meson can be expressed as
\begin{equation}
\frac{\deriv^5 (\Gamma + \overline{\Gamma})}{\deriv {\rm cos}\,\theta_1 \, \deriv {\rm cos}\,\theta_2 \, \deriv\phi \, \deriv m_1 \, \deriv m_2} =
N \sum_{i=1}^6 \sum_{j \geq i}^6 \Real \left[ A_i A_j^* \left( \frac{1-\eta_i}{\GH} + \frac{1 + \eta_i}{\GL} \right) F_{ij} \delta_{\eta_i \eta_j} \right],
\label{eq:rate_timeint}
\end{equation}
where the $A_i$ amplitudes account for the the average of \Bq and \Bqb decays and $N$ is a normalisation constant. For the \Bd meson, a further simplification of the decay rate is considered, since $\Delta \Gamma/\Gamma=-0.002 \pm 0.010$~\cite{HFLAV} the light and heavy mass eigenstate widths can be assumed to be equal,
\[
\left( \frac{1-\eta_i}{\GH} + \frac{1 + \eta_i}{\GL} \right) \approx \frac{2}{\Gamma},
\]
and this factor can be extracted as part of the normalisation constant in~\eqref{eq:rate_timeint}. For the \Bs meson the central values $\GH = 0.618$\invps and $\GL = 0.708$\invps~\cite{HFLAV} are considered.

\section{Detector and simulation}
\label{sec:Detector}

The \lhcb detector~\cite{Alves:2008zz,LHCb-DP-2014-002} is a single-arm forward spectrometer covering the \mbox{pseudorapidity} range $2<\eta <5$, designed for the study of particles containing \bquark or \cquark quarks. The detector includes a high-precision tracking system consisting of a silicon-strip vertex detector surrounding the $pp$ interaction region, a large-area silicon-strip detector located upstream of a dipole magnet with a bending power of about $4{\mathrm{\,Tm}}$, and three stations of silicon-strip detectors and straw drift tubes placed downstream of the magnet. The tracking system provides a measurement of the momentum, \ptot, of charged particles with a relative uncertainty that varies from 0.5\% at low momentum to 1.0\% at 200\gevc. The minimum distance of a track to a primary vertex (PV), the impact parameter (IP), is measured with a resolution of $(15+29/\pt)\mum$, where \pt is the component of the momentum transverse to the beam, in\,\gevc. Different types of charged hadrons are distinguished using information from two ring-imaging Cherenkov detectors. Photons, electrons and hadrons are identified by a calorimeter system consisting of scintillating-pad and preshower detectors, an electromagnetic and a hadronic calorimeter. Muons are identified by a system composed of alternating layers of iron and multiwire proportional chambers.

The magnetic field deflects oppositely charged particles in opposite directions and this can lead to detection asymmetries. Periodically reversing the magnetic field polarity throughout the data-taking almost cancels the effect. The configuration with the magnetic field pointing upwards (downwards), \MagUp (\MagDown), bends positively (negatively) charged particles in the horizontal plane towards the centre of the LHC ring.

The online event selection is performed by a trigger~\cite{LHCb-DP-2012-004}, which consists of a hardware stage, based on information from the calorimeter and muon systems, followed by a software stage, which applies a full event reconstruction. In the offline selection, trigger signatures are associated with reconstructed particles. Since the trigger system uses the \pt of the charged particles, the phase-space and time acceptance is different for events where signal tracks were involved in the trigger decision (called trigger-on-signal or TOS throughout) and those where the trigger decision was made using information from the rest of the event only (noTOS).

Simulated samples of the \BdKstKst and \BsKstKst decays with longitudinal polarisation fractions of 0.81 and 0.64, respectively, are primarily employed in these analyses, particularly for the acceptance description as explained in~\secref{sec:amplitude-analysis}. Simulated samples of the main peaking background contributions, \decay{\Bz}{\Kstarz\phi(\Kp\Km)}, \decay{\Bz}{\rhoz\Kstarz} and \decay{\Lb}{\Kstarz\proton\pim}, are also considered. In the simulation, $pp$ collisions are generated using \pythia~\cite{Sjostrand:2006za} with a specific \lhcb configuration~\cite{LHCb-PROC-2010-056}. Decays of hadronic particles are described by \evtgen~\cite{Lange:2001uf}, in which final-state radiation is generated using \photos~\cite{Golonka:2005pn}. The interaction of the generated particles with the detector, and its response, are implemented using the \geant toolkit~\cite{Allison:2006ve, Agostinelli:2002hh} as described in Ref.~\cite{LHCb-PROC-2011-006}.

\section{Signal selection}
\label{sec:selection}

Both data and simulation are filtered with a preliminary selection. Events containing four good quality tracks with $\pt > 500$\mevc are retained. In events that contain more than one PV, the \Bq candidate constructed with these four tracks is associated with the PV that has the smallest \chisqip, where \chisqip is defined as the difference in the vertex-fit \chisq of the PV reconstructed with and without the track or tracks in question. Each of the four tracks must fulfil $\chisqip>9$ with respect to the PV and originate from a common vertex of good quality (${\chisqndf < 15}$, where ndf is the number of degrees of freedom of the vertex). To identify kaons and pions, a selection in the difference of the log-likelihoods of the kaon and pion hypothesis (\dllkpi) is applied. This selection is complemented with fiducial constraints that optimise the particle identification determination: the pion and kaon candidates are required to have $3 < p < 100$\gevc and $1.5 < \eta < 4.5$ and be inconsistent with muon hypothesis. The final state opposite charge \Kpi pairs are combined into \Kstarz and \Kstarzb candidates with a mass within $150$\mevcc of the \Kstarz mass. The \Kstarz and \Kstarzb candidates must have $\pt > 900$\mevc and vertex $\chisqndf < 9$. The intermediate resonances must combine into \Bq candidates within $500$\mevcc of the \Bs mass, with a distance of closest approach between their trajectories of less than $0.3$\mm. To guarantee that the \Bq candidate originates in the interaction point, the cosine of the angle between the \Bq momentum and the direction of flight from the PV to the decay vertex is required to be larger than 0.99 and the \chisqip with respect to the PV has to be smaller than 25.

A multivariate selection based on a Boosted Decision Tree with Gradient Boost~\cite{Breiman,*Roe} (BDTG) is employed. It relies on the aforementioned variables and on the \Bq candidate flight distance with respect to the PV and its \pt. Simulated \BdKstKst decays with tracks matched to the generator particles and filtered with the preliminary selection are used as signal sample, whereas the four-body invariant-mass sideband $5600 < M\KpiKpi < 5800\mevcc$, composed of purely combinatorial $(\Kp\pim)(\Km\pip)$ combinations, is used as background sample for the BDTG training. The number of events in the signal training sample of the BDTG is determined using the ratio between the \Bs and the \Bd yields from Ref.~\cite{LHCb-PAPER-2014-068} and the \Bs yield obtained with a four-body mass fit to the data sample after the preliminary selection. The number of events in the background training sample of the BDTG is estimated by extrapolating the background yield in the sideband into the $\pm 30$\mevcc window around the \Bd mass. The requirement on the BDTG output is chosen to maximise the figure of merit $N_S/\sqrt{N_S+N_B}$, where $N_S$ and $N_B$ are the expected output signal and background yields, respectively. Different BDTGs are implemented for 2011 and 2012 data.

A comprehensive search for peaking backgrounds, mainly involving intermediate charm particles, is performed. Decays of \B mesons sharing the same final state with the signal,\footnote{The branching fractions in this section are taken from Ref.~\cite{pdg-2018}.} such as ${\decay{\Bd}{\Dz(\Kp \Km) \pip \pim}}$ ($\BR\sim3\times10^{-6}$) and \decay{\Bd}{\Dz(\pip \pim) \Kp \Km} ($\BR\sim6\times10^{-8}$) decays, are strongly suppressed by the requirement in the \Kpi mass. Resonances in three-body combinations (\Kp\Km\pip) and (\Kp\pip\pim) are also explored. In the case of the former, the three-body invariant mass in the data sample is above all known charm resonances. For the latter, no evidence of candidates originated in \decay{\Bd}{\Dmp\Kpm} and \decay{\Bd}{\Dsmp\Kpm} decays ($\BR\sim10^{-7}$) or in \decay{\Bs}{\Dsmp\Kpm} decays ($\BR\sim 10^{-6}$) is found. Three-body combinations with a pion misidentified as a kaon are reconstructed, mainly searching for \decay{\Bz}{\Dm (\pim\pim \Kp) \pip} decays ($\BR = 2.45 \times 10^{-4}$), but also for \decay{\Bs}{\Dspm \Kmp} ($\BR \sim 10^{-5}$), \decay{\Bz}{\Dm\Kp} and \decay{\Bz}{\Dsm\Kp} ($\BR \sim 10^{-6}$) decays. All of them are suppressed to a negligible level by the applied selection. A search of three-body combinations with a proton misidentified as a kaon is performed, finding no relevant contribution from decays involving a \Lc baryon. Decays into five final-state particles are also investigated. Contributions of the \decay{\Bz}{\eta^{\prime}(\gamma\pip\pim) \Kstarz} decay can be neglected due to the small misidentification probability and the four-body mass distribution whereas the \decay{\Bs}{\phi (\pi^0 \pi^+ \pi^-) \phi(K^+ K^-)} decay is negligible due to the requirement on the \Kpi mass.

\section{Four-body mass spectrum}
\label{sec:4body}

The signal and background yields are determined by means of a simultaneous extended maximum-likelihood fit to the invariant-mass spectra of the four final-state particles in the 2011 and 2012 data samples. The \BqKpiKpi signal decays are parameterised with double-sided Hypatia distributions~\cite{MartinezSantos2014150} with the same parameters except for their means that are shifted by the difference between the \Bd and \Bs masses, $87.13\mevcc$~\cite{pdg-2018}. Misidentified \decay{\Bz}{(\Kp \pim)(\Km\Kp)} (including \decay{\Bz}{\Kstarz\phi} decays), \decay{\Lb}{(\proton \pim)(\Kp \pim)}  and \decay{\Bz}{\rhoz \Kstarz} decays are also considered in the fit. Both the \decay{\Bz}{(\Kp \pim)(\Km\Kp)} and \decay{\Lb}{(\proton \pim)(\Km \pip)} contributions are described with the sum of a Crystal Ball~\cite{Skwarnicki:1986xj} function and a Gaussian distribution which shares mean with the Crystal Ball core. The parameters of these distributions are obtained from simulation, apart from the mean and resolution values which are free to vary in the fit. Whereas the distribution mean values are constrained to be the same in the 2011 and 2012 data, the resolution is allowed to have different values for the two samples. The small contributions from \decay{\Bz}{\rhoz \Kstarz} and \decay{\Lb}{(\proton \pim)(\Km \pip)} decays have a broad distribution in the four-body mass and are the object of specific treatment. The contribution from \decay{\Bz}{\rhoz \Kstarz} decays has an expected yield of $3.5 \pm 1.3$ ($6.6 \pm 2.3$) in the 2011 (2012) sample. It is estimated from the detection and selection efficiency measured with simulation, the collected luminosities, the cross section for $\bquark\bquarkbar$ production, the hadronisation fractions of \Bd and \Bs mesons and the known branching fraction of the mode. Simulated events containing this decay mode are added with negative weights to the final data sample to subtract its contribution. The contribution of \decay{\Lb}{(\proton \pim)(\Km \pip)} decays in the 2011 (2012) sample is determined to be $36 \pm 16$ ($120 \pm 28$) from a fit to the $(\proton \pim \Km \pip)$ four-body mass spectrum of the selected data. In this study the four-body invariant mass is recomputed assigning the proton mass to the kaon with the largest \dllpk value. In these fits the \Lb component is described with a Gaussian distribution and the dominant \BsKpiKpi background is described with a Crystal Ball function. The parameters of both lineshapes are obtained from simulation. The remaining contributions, mainly \decay{\Bz}{(\Kp \pim)(\Km\Kp)} and partially reconstructed events, are parameterised with a decreasing exponential with a free decay constant. The \decay{\Lb}{(\proton \pim)(\Km \pip)} decay angular distribution is currently unknown and its contribution can not be subtracted with negatively weighted simulated events. Its subtraction is commented further below.

Finally, contributions from partially reconstructed \bquark-hadron decays and combinatorial background are also considered. The former is composed of \B- and \Bs-meson decays containing neutral particles that are not reconstructed. Because of the missing particle, the measured four-body invariant mass of these candidates lies in the lower sideband of the spectrum. All contributions to this background are jointly parameterised with an ARGUS function~\cite{Albrecht:1990cs} convolved with a Gaussian resolution function, with the same width as the signal. The endpoint of the distribution is also fixed to the \Bs mass minus the \piz mass. The combinatorial background is composed of charged tracks that are not originating from the signal decay chain. It is modelled with a linear distribution, with a free slope parameter, separate for 2011 and 2012 data samples.

\begin{figure}[t]
\begin{center}
\includegraphics[width=0.7\textwidth]{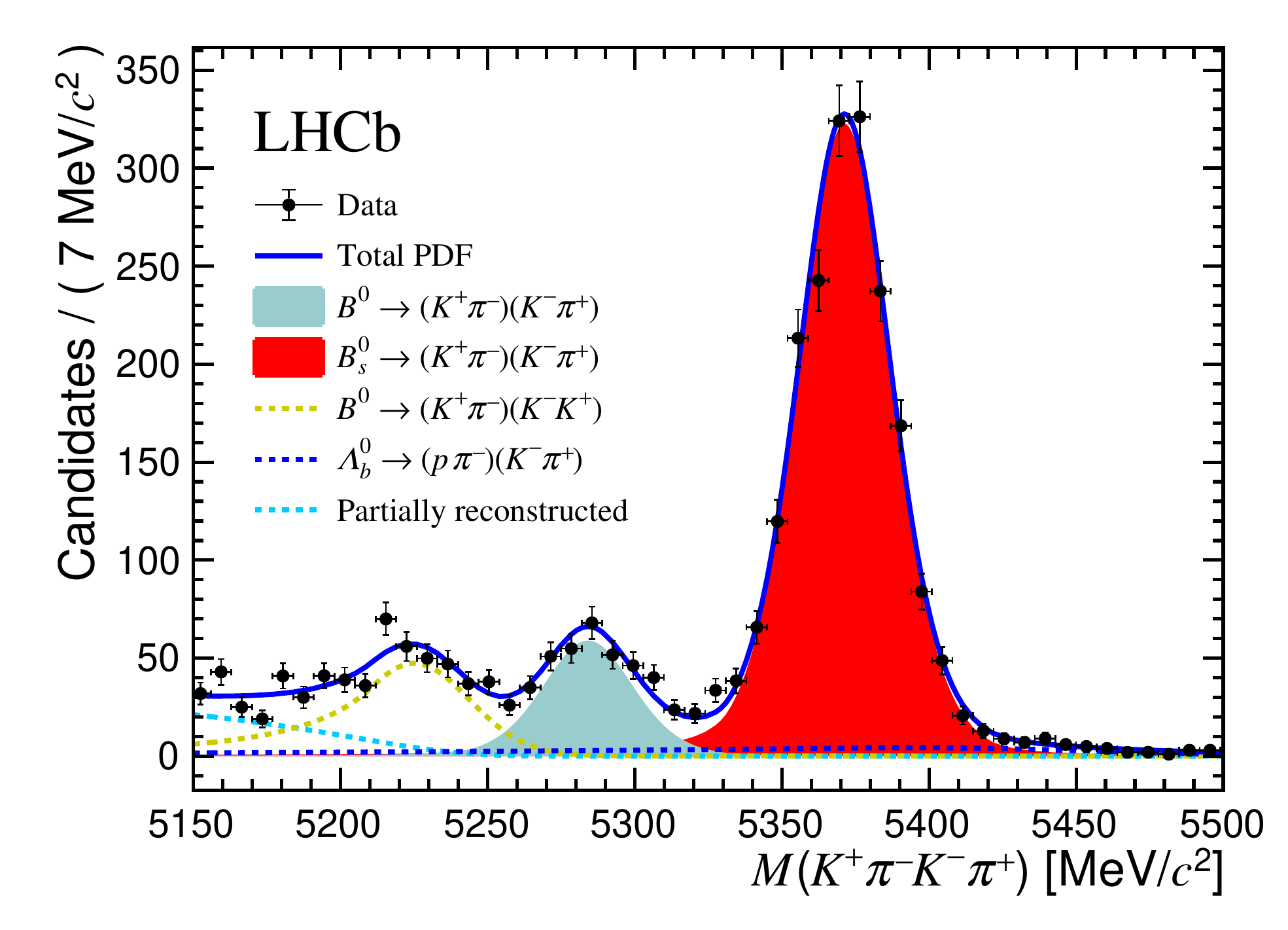}
\caption{Aggregated four-body invariant-mass fit result of the 2011 and 2012 data. The solid red distribution corresponds to the \BsKpiKpi decay, the solid cyan distribution to \BdKpiKpi, the dotted dark blue line to \decay{\Lb}{(\proton \pim)(\Km \pip)}, the dotted yellow line to \decay{\Bz}{(\Kp \pim)(\Km\Kp)} and the dotted cyan line represents the partially reconstructed background. The tiny combinatorial background contribution is not represented. The black points with error bars correspond to data to which the \decay{\Bz}{\rhoz \Kstarz} contribution has been subtracted with negatively weighted simulation, and the overall fit is represented by the thick blue line.\label{fig:4body-fit}}
%(For interpretation of the references to colour in this figure caption, the reader is referred to the web version of this paper.)}
\end{center}
\end{figure}

The results of the fit to the four-body mass spectrum are shown in~\figref{fig:4body-fit} and the yields are reported in~\tabref{tab:4body-result}. In total, about three hundred \BdKpiKpi signal candidates are found, a factor seven larger than previous analyses~\cite{babar,Chiang:2010ga}. To perform a background-subtracted amplitude analysis, the \sPlot technique~\cite{Pivk:2004ty,Xie:2009rka} is applied to isolate either the \BdKpiKpi or the \BsKpiKpi decays. The contribution from \decay{\Lb}{(\proton \pim)(\Km \pip)}, for which the yield is fixed, is treated using extended weights according to Appendix B.2 of Ref.~\cite{Pivk:2004ty}. The \sPlot method suppresses the background contributions using their relative abundance in the four-body invariant mass spectrum and, therefore, no assumption is required for their phase-space distribution.  

\begin{table}[t]
\caption{Signal and background yields for the 2011 and 2012 data samples, obtained from the fit to the four-body mass spectrum of the selected candidates. Statistical and systematic uncertainties are reported, the latter are estimated as explained in~\secref{sec:branching}.}
\label{tab:4body-result}
\begin{center}
\begin{tabular}{l c  c }
Yield & 2011 sample & 2012 sample \\
\hline
\decay{\Bd}{(\Kp \pim)(\Km \pip)} & $\phantom{0}99 \pm 12 \pm \phantom{0}3$ & $\phantom{0}249 \pm 19 \pm \phantom{0}5$ \\
\decay{\Bs}{(\Kp \pim)(\Km \pip)} & $617 \pm 26 \pm \phantom{0}8$ & $1337 \pm 39 \pm 12$ \\
Misidentified \decay{\Bd}{(\Kp \pim)(\Km\Kp)} & $145 \pm 17 \pm \phantom{0}2$ & $\phantom{0}266 \pm 27 \pm \phantom{0}8$ \\
Partially reconstructed background & $100 \pm 15 \pm \phantom{0}4$ & $\phantom{0}230 \pm 25 \pm \phantom{0}6$ \\
Combinatorial background & $\phantom{00}7 \pm \phantom{0}5 \pm 11$ & $\phantom{00}48 \pm 25 \pm 25$ \\
\end{tabular}
\end{center}
\end{table}

\section{Amplitude analysis}
\label{sec:amplitude-analysis}

Each of the background-subtracted samples of \BdKpiKpi and \BsKpiKpi decays is the object of a separate amplitude analysis based on the model described in~\secref{sec:amplitude-formalism}. As a first step, the effect of a non-uniform efficiency, depending on the helicity angles and the two-body invariant masses, is examined. For this purpose, four categories are defined according to the hardware trigger decisions (TOS or noTOS) and data-taking period (2011 and 2012). The efficiency is accounted for through the complex integrals~\cite{TristansThesis}
\begin{equation}
\label{equation:n_weights}
\omega^k_{ij} =
\int\varepsilon(m_1,m_2,\theta_1,\theta_2,\phi)F_{ij}\deriv m_1 \deriv m_2 \deriv{\rm cos}\,\theta_1 \deriv {\rm cos}\,\theta_2 \deriv\phi,
\end{equation}
\noindent where $\varepsilon$ is the total phase-space dependent efficiency, $k$ is the sample category and $F_{ij}$ are defined in~\eqref{eq:Fij}. The integrals of \eqref{equation:n_weights} are determined using simulated signal samples of each of the four categories, selected with the same criteria applied to data. A single set of integrals is used for both the \Bs and the \Bd amplitude analyses. A probability density function (\PDF) for each category is built 
\begin{equation}
\label{eq:pdfwithacc}
S^k (m_1,m_2,\theta_1,\theta_2,\phi) = \frac{\sum\limits_{i=1}^6 \sum\limits_{j \geq i}^6 \Real \left[ A_i A_j^* \left( \frac{1-\eta_i}{\GH} + \frac{1 + \eta_i}{\GL} \right) F_{ij} \delta_{\eta_i \eta_j} \right]}{\sum\limits_{i=1}^6 \sum\limits_{j \geq i}^6 \Real\left[A_iA_j^* \left( \frac{1-\eta_i}{\GH} + \frac{1 + \eta_i}{\GL} \right)\omega_{ij}^k \delta_{\eta_i\eta_j}\right]},
\end{equation}
where $A_i$ and $\eta_i$ are given in~\tabref{tab:amplitudes}.

Candidates from all categories are processed in a simultaneous unbinned maximum-likelihood fit, separately for each signal decay mode, using the PDFs in~\eqref{eq:pdfwithacc}. To avoid nonphysical values of the parameters during the minimisation, some of them are redefined as
\begin{align}
\begin{split}
f_\parallel &= x_{f_\parallel} (1 - f_L), \\
f_\perp &= (1-x_{f_\parallel})(1 - f_L), \\
|A_S^+|^2 &= x_{|A_S^+|^2}(1 - |A_S^-|^2), \\
|A_{SS}|^2 &= x_{|A_{SS}|^2}(1 - |A_S^-|^2 -|A_S^+|^2),
\nonumber
\end{split}
\end{align}
where $x_{f_\parallel}$, $x_{|A_S^+|^2}$ and $x_{|A_{SS}|^2}$ are used in the fit, together with $f_{L}$, $|A_S^-|^2$,$\delta_\parallel$, $\delta_\perp - \delta_S^+$, $\delta_S^-$ and $\delta_{SS}$. The former three variables are free to vary within the range $[0,1]$, ensuring that the sum of all the squared amplitudes is never greater than $1$. The fit results are corrected for a small reducible bias, originated in discrepancies between data and simulation, as explained in~\secref{sec:systematic-uncertainties}. The final results are shown in~\tabref{tab:amplitude-result}.

\begin{table}[t]
\caption{Results of the amplitude analysis of \BdKpiKpi and \BsKpiKpi decays. The observables above the line are directly obtained from the maximum-likelihood fit whereas those below are obtained from the former, as explained in the text, with correlations accounted for in their estimated uncertainties. For each result, the first quoted uncertainty is statistical and the second systematic. The estimation of the latter is described in~\secref{sec:systematic-uncertainties}.}
\label{tab:amplitude-result}
\begin{center}
\begin{tabular}{ c c c}
Parameter & \BdKstKst  & \BsKstKst \\
\hline
$f_{L}$  &   $0.724 \pm 0.051 \pm 0.016 $  & $0.240 \pm 0.031 \pm 0.025$ \\
$x_{f_\parallel}$  &   $\phantom{0}0.42 \pm \phantom{0}0.10 \pm \phantom{0}0.03$ & $0.307 \pm 0.031 \pm 0.010$ \\
$|A_S^-|^2$  &   $0.377 \pm 0.052 \pm 0.024 $  & $0.558 \pm 0.021 \pm 0.014$ \\
$x_{|A_S^+|^2}$  &   $ 0.013 \pm 0.027 \pm 0.011$ & $0.109 \pm 0.028 \pm 0.024$ \\
$x_{|A_{SS}|^2}$  &   $ 0.038 \pm 0.022 \pm 0.006$ & $0.222 \pm 0.025 \pm 0.031$ \\
$\delta_\parallel$  &   $ \phantom{0}2.51 \pm \phantom{0}0.22 \pm \phantom{0}0.06$ & $\phantom{0}2.37 \pm \phantom{0}0.12 \pm \phantom{0}0.06$ \\
$\delta_\perp - \delta_S^+$  &   $\phantom{0}5.44 \pm \phantom{0}0.86 \pm \phantom{0}0.22$ & $\phantom{0}4.40 \pm \phantom{0}0.17 \pm \phantom{0}0.07$ \\
$\delta_S^-$  &   $\phantom{0}5.11 \pm \phantom{0}0.13 \pm \phantom{0}0.04$  & $\phantom{0}1.80 \pm \phantom{0}0.10 \pm \phantom{0}0.06$ \\
$\delta_{SS}$  &   $\phantom{0}2.88 \pm \phantom{0}0.35 \pm \phantom{0}0.13$ & $\phantom{0}0.99 \pm \phantom{0}0.13 \pm \phantom{0}0.06$ \\
\hline
$f_{\parallel}$   &   $ 0.116 \pm 0.033 \pm 0.012$ & $0.234 \pm 0.025 \pm 0.010$ \\
$f_{\perp}$   &   $ 0.160 \pm 0.044 \pm 0.012$ & $0.526 \pm 0.032 \pm 0.019$ \\
$|A_S^+|^2$   &   $ 0.008 \pm 0.013 \pm 0.007$ & $0.048 \pm 0.014 \pm 0.011$ \\
$|A_{SS}|^2$   &   $ 0.023 \pm 0.014 \pm 0.004$ & $0.087 \pm 0.011 \pm 0.011$ \\
\swave fraction & $0.408 \pm 0.050 \pm 0.017$ & $0.694 \pm 0.016 \pm 0.010$ \\ 
\end{tabular}
\end{center}
\end{table}

\begin{figure}[t]
\begin{center}
\includegraphics[width=\textwidth]{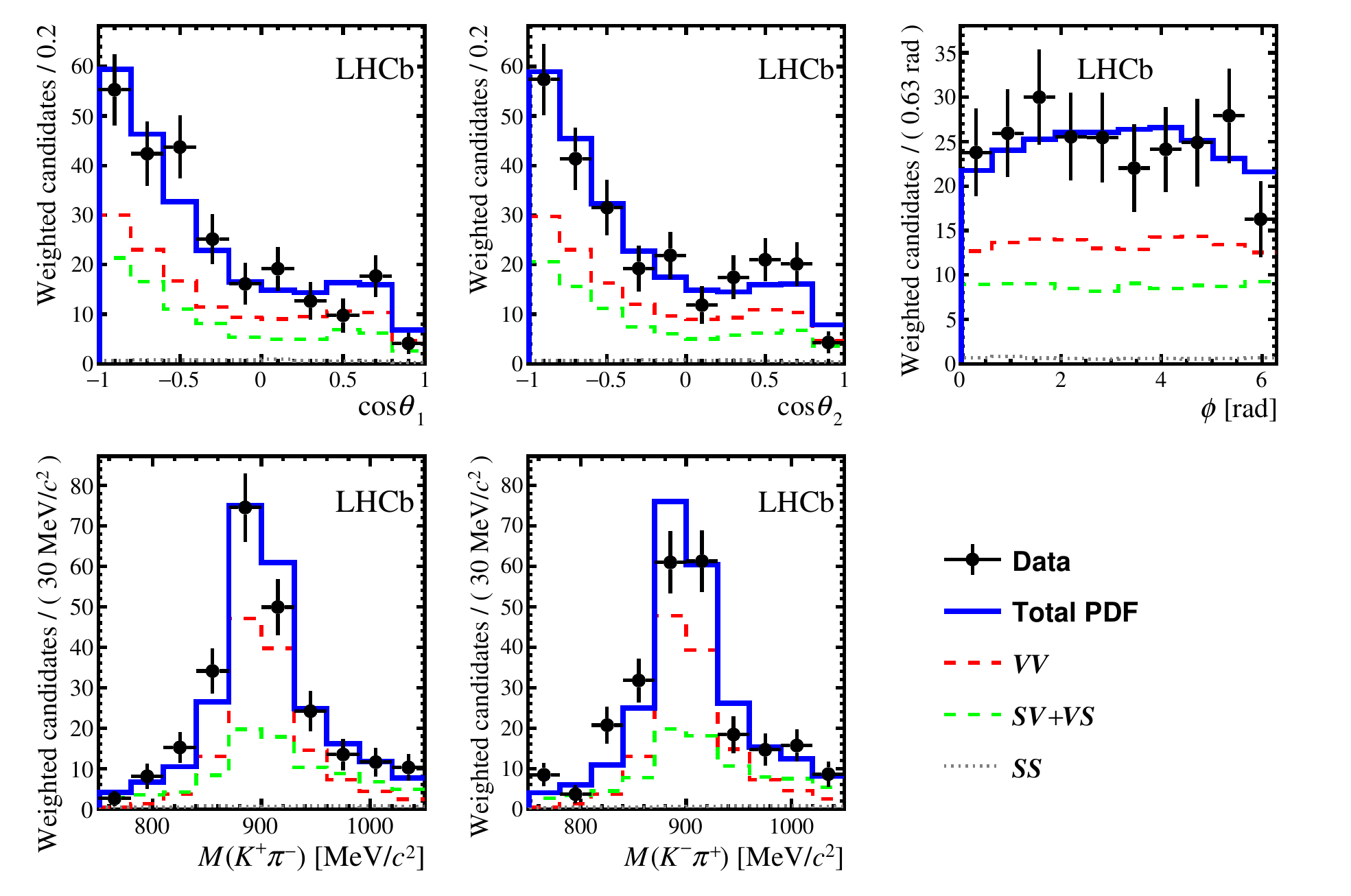}
\end{center}
\caption{Projections of the amplitude fit results for the \BdKstKst decay mode on the helicity angles (top row: $\cos \theta_1$ left, $\cos \theta_2$ centre and $\phi$ right) and on the two-body invariant masses (bottom row: $M(\Kp\pim)$ left and $M(\Km\pip)$ centre). The contributing partial waves: $VV$ (dashed red), $VS$ (dashed green) and $SS$ (dotted grey) are shown with lines. The black points correspond to data and the overall fit is represented by the blue line.\label{fig:mc_visualization_Bd}}
%(For interpretation of the references to colour in this figure caption, the reader is referred to the web version of this paper.)}
\end{figure}

\begin{figure}[ht]
\begin{center}
\includegraphics[width=\textwidth]{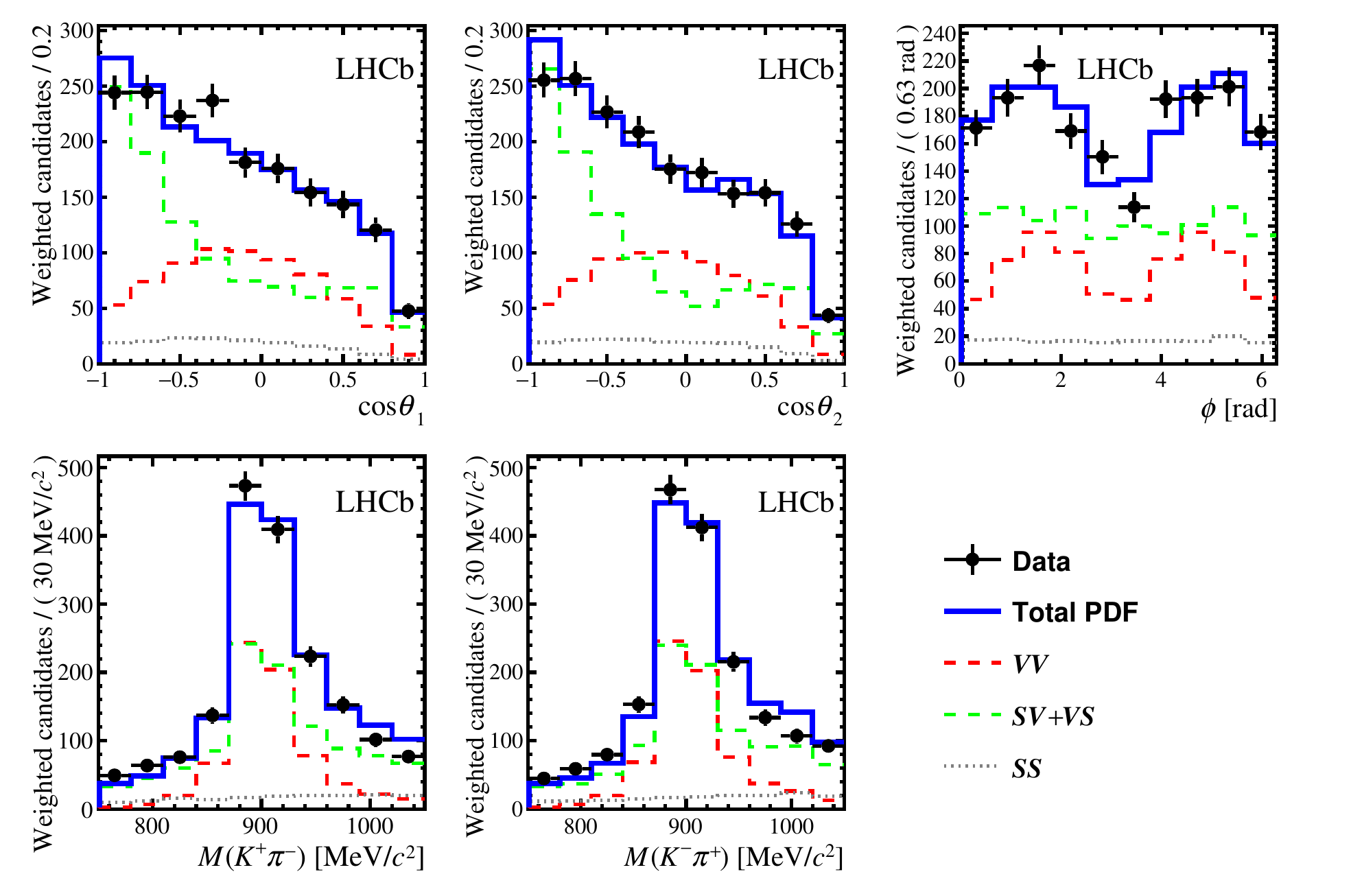}
\end{center}
\caption{Projections of the amplitude fit results for the \BsKstKst decay mode on the helicity angles (top row: $\cos \theta_1$ left, $\cos \theta_2$ centre and $\phi$ right) and on the two-body invariant masses (bottom row: $M(\Kp\pim)$ left and $M(\Km\pip)$ centre). The contributing partial waves: $VV$ (dashed red), $VS$ (dashed green) and $SS$ (dotted grey) are shown with lines. The black points correspond to data and the overall fit is represented by the blue line.\label{fig:mc_visualization_Bs}}
%(For interpretation of the references to colour in this figure caption, the reader is referred to the web version of this paper.)}
\end{figure}

Figs.~\ref{fig:mc_visualization_Bd} and~\ref{fig:mc_visualization_Bs} show the one-dimensional projections of the amplitude fit to the \BdKpiKpi and \BsKpiKpi signal samples in which the background is statistically subtracted by means of the \sPlot technique. Three contributions are shown: $VV$, produced by \Kpia\Kpib pairs originating in a \Kstarz \Kstarzb decay; $VS$, accounting for amplitudes in which only one of the \Kpi pairs originates in a \Kstarz decay; and $SS$, where none of the two \Kpi pairs originate in a \Kstarz decay.

The fraction of $VV$ decays, or purity at production, of the \BdKstKst signal, $f^P_{\Bd}$, is estimated from the amplitude analysis and found to be
\[
f^P_{\Bd} \equiv 1-|A_{SS}|^2-|A_{S}^{+}|^2-|A_{S}^{-}|^2= 0.592 \pm 0.050 \stat \pm 0.017 \syst.
\]
The significance of this magnitude, computed as its value over the sum in quadrature of the statistical and systematic uncertainty, is found to be 10.8 standard deviations. This significance corresponds to the presence of \BdKstKst $VV$ decays in the data sample. The \swave fraction of the decay is equal to $0.408=1-f^P_{\Bd}$. For the \BsKstKst mode the \swave fraction is found to be ${0.694\pm 0.016 \stat \pm 0.010 \syst}$.

\section{Systematic uncertainties of the amplitude analysis}
\label{sec:systematic-uncertainties}

Several sources of systematic uncertainty that affect the results of the amplitude analysis are considered and discussed in the following.

\begin{description}
\item {\bf Fit method.} 
Biases induced by the fitting method are evaluated with a large ensemble of pseudoexperiments. For each signal decay, samples with the same yield of signal observed in data (see ~\tabref{tab:4body-result}) are generated according to the PDF of~\eqref{eq:rate_timeint} with inputs set to the results summarised in~\tabref{tab:amplitude-result}. The use of the weights defined in~\eqref{equation:n_weights} to account the detector acceptance would require a full simulation and, instead, a parametric efficiency is considered. For each observable, the mean deviation of the result from the input value is assigned as a systematic uncertainty.

\item {\bf Description of the kinematic acceptance.}
The uncertainty on the signal efficiency relies on the coefficients of~\eqref{equation:n_weights} that are estimated with simulation. To evaluate its impact on the amplitude analysis results, the fit to data is repeated several times with alternative coefficients varied according to their covariance matrix. The standard deviation of the distribution of the fit results for each observable is assigned as a systematic uncertainty.

\item {\bf Resolution.} The fit performed assumes a perfect resolution on the phase-space variables. The impact of the detector resolution on these variables is estimated with sets of pseudoexperiments adding per-event random deviations according to the resolution estimated from simulation. For each observable, the mean deviation of the result from the measured value is assigned as a systematic uncertainty.

\item {\bf \pwave mass model.} The amplitude analysis is repeated with alternative values of the parameters that define the \pwave mass propagator, detailed in~\tabref{table:inputSwave}, randomly sampled from their known values~\cite{pdg-2018}. The standard deviation of the distribution of the amplitude fit results for each observable is assigned as a systematic uncertainty.

\item {\bf \swave mass model.} In addition to the default \swave propagator, described in~\secref{sec:amplitude-formalism}, two alternative models are used: the LASS lineshape with the parameters of~\tabref{tab:mass-values-jgp}, obtained with \mbox{\decay{\Bd}{J/\psi \Kp \pim}} decays within the analysis of Ref.~\cite{LHCb-PAPER-2014-014}, and the propagator proposed in Ref.~\cite{Pelaez:2016tgi}. The amplitude fit is performed with these two alternatives and, for each observable, the largest deviation from the baseline result is assigned as a systematic uncertainty.

\begin{table}
\caption{Alternative parameters of the LASS mass propagator used in the \swave systematic uncertainty estimation.}
\label{tab:mass-values-jgp}
\begin{center}
\begin{tabular}{@{}c@{}@{}l@{}|@{}r@{}@{}c@{}@{}l@{}}
  &   & \multicolumn{3}{c}{$(\Kp \pim)_0$}  \\
\hline
$M_0$ & \phantom{,}[MeV/$c^2$]\phantom{,,} & $\, 1456.7\,$ & $\pm$ & $\,3.9$ \\
$\Gamma_0$ & \phantom{,}[MeV] & $323\,$ & $\pm$ & $\,11$ \\
$a$  & \phantom{,}[$c$/GeV] & $3.83\,$ & $\pm$ & $\,0.11$ \\
$b$ & \phantom{,}[$c$/GeV]  & $2.86\,$ & $\pm$ & $\,0.22$ \\
\end{tabular}
\end{center}
\end{table}

\item {\bf Differences between data and simulation.} An iterative method~\cite{GarciaPardinas:2630181}, is used to weight the simulated events and improve the description of the track multiplicity and \Bq-meson momentum distributions. The procedure is repeated multiple times and, for each observable, the mean bias of the amplitude fit result is corrected for in the results of~\tabref{tab:amplitude-result} while its standard deviation is assigned as a systematic uncertainty. 

\item {\bf Background subtraction.} The data set used in the amplitude analysis is background subtracted using the \sPlot method~\cite{Pivk:2004ty,Xie:2009rka} that relies in the lineshapes of the four-body mass fit discussed in~\secref{sec:4body}. The uncertainty related to the determination of the signal weights is evaluated repeating the amplitude analysis fits with weights obtained fitting the four-body invariant-mass with two alternative models. In the first case, the model describing the signal is defined by the sum of two Crystal Ball functions~\cite{Skwarnicki:1986xj} with a common, free, peak value and the resolution parameter fixed from simulation. In the second case, the model describing the combinatorial background is assumed to be an exponential function. The amplitude fit is performed with the \sPlot-weights obtained with the two alternatives and, for each observable, the largest deviation from the baseline result is assigned as a systematic uncertainty. This procedure is also used when addressing the systematic uncertainties in the measured yields of the different subsamples, as discussed in~\secref{sec:branching}.

\item {\bf Peaking backgrounds.} The uncertainty related to the fluctuations in the yields of the \decay{\Lb}{(p \pim)(\Km \pip)} and \decay{\Bd}{\rhoz \Kstarz} background contributions are estimated repeating the amplitude-analysis fit with the yield values varied by their uncertainties reported in~\secref{sec:4body}. For each observable, the largest deviation from the default result is assigned as a systematic uncertainty. This procedure is also used when addressing the systematic uncertainties of the four-body invariant mass yields in~\secref{sec:branching}.

\item {\bf Time acceptance.} The amplitude analysis does not account for possible decay-time dependency of the efficiency, however, the trigger and the offline selections may have an impact on it. This effect only affects \Bs-meson decays and is accounted for by estimating effective shifts: $\GH = 0.618 \to 0.598$\invps and $\GL = 0.708 \to 0.732$\invps, which are obtained with simulation. For each observable, the variation of the result of the fit after introducing these values in the amplitude analysis is considered as a systematic uncertainty.
\end{description}

The resulting systematic uncertainties and the corrected biases, originated in the differences between data and simulation, are detailed in~\tabref{tab:sistBd} for the parameters of the amplitude-analysis fit. The corresponding values for the derived observables are summarised in~\tabref{tab:sistBd_derived}. The total systematic uncertainty is computed as the sum in quadrature of the different contributions.

\begin{table}[t]
\caption{Systematic uncertainties for the parameters of the amplitude-analysis fit of the \BqKpiKpi decay. The bias related to differences between data and simulation is included in the results shown in~\tabref{tab:amplitude-result}.\label{tab:sistBd}}
\begin{center}
\resizebox{\textwidth}{!}{
\begin{tabular}{l|c c c c c c c c c }
Decay mode & \multicolumn{9}{c}{\BdKpiKpi} \\ 
\hline
Parameter &$f_{L}$&$x_{f_\parallel}$&$|A_S^-|^2$&$\phantom{-}x_{|A_S^+|^2}$&$x_{|A_{SS}|^2}$&$\delta_\parallel$&$\delta_\perp - \delta_S^+$&$\delta_S^-$&$\delta_{SS}$ \\
\hline
Bias data-simulation & 0.001 & 0.00 & 0.006 & $-0.001$ & 0.004 & 0.01 & $-0.01$ & 0.00 & 0.01 \\ 
\hline
Fit method & 0.007 & 0.01 & 0.011 & $\phantom{-}0.009$ & 0.001 & 0.00 & $\phantom{-}0.01$ & 0.00 & 0.02 \\
Kinematic acceptance & 0.005 & 0.01 & 0.006 & $\phantom{-}0.004$ & 0.002 & 0.03 & $\phantom{-}0.12$ & 0.01 & 0.04 \\
Resolution & 0.007 & 0.00 & 0.005 & $\phantom{-}0.001$ & 0.002 & 0.00 & $\phantom{-}0.16$ & 0.00 & 0.02 \\
\pwave mass model & 0.001 & 0.00 & 0.004 & $\phantom{-}0.001$ & 0.002 & 0.00 & $\phantom{-}0.01$ & 0.00 & 0.02 \\
\swave mass model & 0.007 & 0.01 & 0.016 & $\phantom{-}0.003$ & 0.002 & 0.03 & $\phantom{-}0.03$ & 0.03 & 0.02 \\
Differences data-simulation & 0.004 & 0.00  & 0.002 & $\phantom{-}0.001$ & 0.001 & 0.01  & $\phantom{-}0.01$  & 0.01  & 0.01 \\
Background subtraction & 0.002 & 0.01  & 0.006 & $\phantom{-}0.001$ & 0.002 & 0.01  & $\phantom{-}0.06$  &0.01  &0.09 \\
Peaking backgrounds & 0.009 & 0.02 & 0.009 & $\phantom{-}0.003$ & 0.003 & 0.04 & $\phantom{-}0.06$ & 0.01 & 0.08 \\
\cdashline{1-10}
Total systematic unc. & 0.016 & 0.03 & 0.024 & $\phantom{-}0.011$ & 0.006 & 0.06 & $\phantom{-}0.22$ & 0.04 & 0.13 \\
\end{tabular}
}
\resizebox{\textwidth}{!}{
\begin{tabular}{l|c c c c c c c c c }
\multicolumn{10}{c}{} \\ 
Decay mode & \multicolumn{9}{c}{\BsKpiKpi} \\      
 \hline
Parameter &$f_{L}$&$x_{f_\parallel}$&$|A_S^-|^2$&$\phantom{-}x_{|A_S^+|^2}$&$x_{|A_{SS}|^2}$&$\delta_\parallel$&$\delta_\perp - \delta_S^+$&$\delta_S^-$&$\delta_{SS}$ \\
\hline
Bias data-simulation & 0.004 & 0.003 & 0.007 & $-0.003$ & 0.021 & 0.05 & $\phantom{-}0.00$ & 0.05 & 0.07 \\
\hline
Fit method & 0.001 & 0.000 & 0.001 & $\phantom{-}0.000$ & 0.000 & 0.00 & $\phantom{-}0.00$ & 0.00 & 0.00 \\
Kinematic acceptance & 0.011 & 0.006 & 0.011 & $\phantom{-}0.021$ & 0.009 & 0.05 & $\phantom{-}0.07$ & 0.05 & 0.05 \\
Resolution & 0.002 & 0.001 & 0.000 & $\phantom{-}0.002$ & 0.000 & 0.00 & $\phantom{-}0.00$ & 0.00 & 0.00 \\
\pwave mass model & 0.001 & 0.000 & 0.001 & $\phantom{-}0.002$ & 0.009 & 0.00 & $\phantom{-}0.01$ & 0.00 & 0.01 \\
\swave mass model & 0.021 & 0.001 & 0.007 & $\phantom{-}0.011$ & 0.028 & 0.03 & $\phantom{-}0.02$ & 0.03 & 0.02 \\
Differences data-simulation & 0.002 & 0.000 & 0.001 & $\phantom{-}0.001$ & 0.001 & 0.01  & $\phantom{-}0.00$  & 0.01  & 0.01 \\
Background subtraction & 0.000 & 0.001  & 0.001 & $\phantom{-}0.001$ & 0.004 & 0.01  & $\phantom{-}0.01$  &0.01  &0.01 \\
Peaking backgrounds & 0.003 & 0.008 & 0.002 & $\phantom{-}0.002$ & 0.002 & 0.02 & $\phantom{-}0.01$ & 0.02 & 0.01 \\
Time acceptance & 0.008 & 0.014 & 0.008 & $\phantom{-}0.004$ & 0.005 & 0.00 & $\phantom{-}0.00$ & 0.00 & 0.00 \\
\cdashline{1-10}
Total systematic unc. & 0.025 & 0.010 & 0.014 & $\phantom{-}0.024$ & 0.031 & 0.06 & $\phantom{-}0.07$ & 0.06 & 0.05 \\
\end{tabular}
}
\end{center}
\end{table}

\begin{table}[t]
\caption{Systematic uncertainties for the derived observables  of the amplitude-analysis fit of the \BqKpiKpi decay. The bias related to differences between data and simulation is included in the results shown in~\tabref{tab:amplitude-result}.\label{tab:sistBd_derived}}
\begin{center}
{
\begin{tabular}{l|c c c c c c c c c }
Decay mode & \multicolumn{9}{c}{\BdKpiKpi} \\      
 \hline
Observable &$f_{\parallel}$&$f_{\perp}$&$|A_S^+|^2$&$|A_{SS}|^2$&\swave fraction \\
\hline
Bias data-simulation & 0.001 & $-0.001$ & $-0.001$ & 0.002& 0.007 \\
\hline
Fit method & 0.000 & $\phantom{-}0.007$ & $\phantom{-}0.005$ & 0.000 & 0.006 \\
Kinematic acceptance & 0.003 & $\phantom{-}0.004$ & $\phantom{-}0.001$ & 0.003 & 0.006 \\
Resolution & 0.001 & $\phantom{-}0.003$ & $\phantom{-}0.000$ & 0.001 & 0.006 \\
\pwave mass model & 0.000 & $\phantom{-}0.001$ & $\phantom{-}0.000$ & 0.001 & 0.005 \\
\swave mass model & 0.000 & $\phantom{-}0.007$ & $\phantom{-}0.002$ & 0.002 & 0.008 \\
Differences data-simulation & 0.001 & $\phantom{-}0.003$ & $\phantom{-}0.000$ & 0.001 & 0.002 \\
Background subtraction & 0.005 & $\phantom{-}0.003$  & $\phantom{-}0.001$ & 0.001 & 0.002 \\
Peaking backgrounds & 0.010 & $\phantom{-}0.003$ & $\phantom{-}0.002$ & 0.002 & 0.009 \\
\cdashline{1-10}
Total systematic unc. & 0.012 & $\phantom{-}0.012$ & $\phantom{-}0.007$ & 0.004 & 0.017 \\
\end{tabular}
}
{
\begin{tabular}{l|c c c c c c c c c }
\multicolumn{10}{c}{} \\ 
Decay mode & \multicolumn{9}{c}{\BsKpiKpi} \\      
 \hline
Observable &$f_{\parallel}$&$f_{\perp}$&$|A_S^+|^2$&$|A_{SS}|^2$&\swave fraction \\
\hline
Bias data-simulation  & 0.001 & $-0.005$ & $-0.002$ & 0.007 & 0.012 \\
\hline
Fit method & 0.001 & $\phantom{-}0.001$ & $\phantom{-}0.000$ & 0.001 & 0.001 \\
Kinematic acceptance & 0.005 & $\phantom{-}0.009$ & $\phantom{-}0.010$ & 0.004 & 0.004 \\
Resolution & 0.000 & $\phantom{-}0.002$ & $\phantom{-}0.000$ & 0.001 & 0.002 \\
\pwave mass model & 0.000 & $\phantom{-}0.001$ & $\phantom{-}0.001$ & 0.003 & 0.005 \\
\swave mass model & 0.006 & $\phantom{-}0.016$ & $\phantom{-}0.004$ & 0.009 & 0.006 \\
Differences data-simulation & 0.001 & $\phantom{-}0.001$ & $\phantom{-}0.000$ & 0.001 & 0.001 \\
Background subtraction & 0.001 & $\phantom{-}0.001$  & $\phantom{-}0.001$ & 0.002 & 0.002 \\
Peaking backgrounds & 0.007 & $\phantom{-}0.005$ & $\phantom{-}0.001$ & 0.001 & 0.001  \\
Time acceptance & 0.008 & $\phantom{-}0.016$ & $\phantom{-}0.003$ & 0.001 & 0.007  \\
\cdashline{1-10}
Total systematic unc. & 0.010 & $\phantom{-}0.019$ & $\phantom{-}0.011$ & 0.011 & 0.010 \\
\end{tabular}
}
\end{center}
\end{table}

\section{Determination of the ratio of branching fractions}
\label{sec:branching}

In this analysis, the \BdKstKst branching fraction is measured relative to that of \BsKstKst decays. Since both decays are selected in the same data sample and share a common final state most systematic effects cancel. However, some efficiency corrections, eg. those originated from the difference in phase-space distributions of events of the two modes, need to be accounted for. The amplitude fit provides the relevant information to tackle the differences between the two decays.

This branching-fraction ratio is obtained as
\begin{equation}
\label{eq:branching-ratio}
\frac{\BF(\BdKstKst)}{\BF (\BsKstKst)}
= \frac{\varepsilon_{\Bs}}{\varepsilon_{\Bd}}
\times \frac{\lambda^{f_L}_{\Bs}}{\lambda^{f_L}_{\Bd}}
\times \frac{N_\Bd \times f^D_{\Bd}}{N_\Bs \times f^D_{\Bs}}
\times \frac{f_s}{f_d},
\end{equation}
where, for each channel, $\varepsilon_{\Bq}$ is the detection efficiency, $\lambda^{f_L}_{\Bq}$ is a polarisation-dependent correction of the efficiency, originated in differences between the measured polarisation and that assumed in simulation, $N_\Bq$ is the measured number of \BqKpiKpi candidates and $f^D_{\Bq}$ represents the $VV$ signal purity at detection. In this way $N_\Bq \times f^D_{\Bq}$ represents the \BqKstKst yield. Finally, $f_d$ and $f_s$ are the hadronisation fractions of a $\bquark$-quark into a \Bd and \Bs meson, respectively.

The purity at detection and the $\lambda^{f_L}$ factor ratios, $\kappa^k_{\Bq}$,  are obtained for each decay mode as
\begin{equation}
\kappa^k_{\Bq} \equiv \frac{\lambda^{f_L}_{\Bq}}{f^D_{\Bq}} =
\frac{\sum\limits_{i=1}^6 \sum\limits_{j \geq i}^6 \Real [A_i A_j^* \left( \frac{1-\eta_i}{\GH} + \frac{1 + \eta_i}{\GL} \right) \omega^k_{ij}]}{(1- |A_S^-|^2 - |A_S^+|^2 - |A_{SS}|^2) \sum\limits_{i=1}^3 \sum\limits_{j \geq i}^3 \Real [A_i^{\rm sim} A_j^{{\rm sim}*} \left( \frac{1-\eta_i}{\GH} + \frac{1 + \eta_i}{\GL} \right) \omega^k_{ij}]}, \label{eq:kappa}
\end{equation}
where the $\omega^k_{ij}$ coefficients are defined in~\eqref{equation:n_weights}, $A_i^{\rm sim}$ are the amplitudes used to generate signal samples, and the $\eta_i$ values are given in~\tabref{tab:amplitudes}. Also in this case, for the \BdKstKst decay, the $\GH=\GL$ approximation is adopted.

The detection efficiency is determined from simulation for each channel separately for the different categories discussed in~\secref{sec:amplitude-analysis}: year of data taking, trigger type and, in addition, the \lhcb magnet polarity. An exception is applied to the particle-identification selection whose efficiency is determined from large control samples of $\Dstarp \to \Dz \pip$, $\Dz \to \Km\pip$ decays. Differences in kinematics and detector occupancy between the control samples and the signal data are accounted for in this particle-identification efficiency study~\cite{LHCb-DP-2012-003,LHCb-PUB-2016-021}.

The different sources of systematic uncertainty in the branching fraction determination are discussed below.
\begin{description}

\item {\bf Systematic uncertainties in the factor $\boldsymbol{\kappa}$.} The uncertainties on the parameters of the amplitude analysis fit described in~\secref{sec:systematic-uncertainties} affect the determination of the factors $\kappa$ defined in~\eqref{eq:kappa} as summarised in~\tabref{tab:sistkappa}.

\item {\bf Systematic uncertainties in the signal yields.} As discussed in~\secref{sec:systematic-uncertainties} uncertainties on the signal yields arise from the model used to fit the four-body invariant mass. The uncertainties from the different proposed alternative signal and background lineshapes are summed in quadrature to compute the final systematic uncertainty.

\item {\bf Systematic uncertainty in the efficiencies.} A dedicated data method is employed to estimate the uncertainty in the signal efficiency originated in the PID selection.

\end{description}

\begin{table}[t]
\caption{Systematic uncertainties in the factor $\kappa$ defined in~\eqref{eq:kappa} split in categories. The bias originated in differences between data and simulation is corrected for in the $\kappa$ results shown in~\tabref{tab:branching-ratio}.\label{tab:sistkappa}}
\begin{center}
\resizebox{\textwidth}{!}{
\begin{tabular}{l|c c| c c | c c |c c }
Decay mode &\multicolumn{4}{c|}{\BdKpiKpi}&\multicolumn{4}{c}{\BsKpiKpi}\\
\hline
Year &\multicolumn{2}{c|}{2011}&\multicolumn{2}{c|}{2012}&\multicolumn{2}{c|}{2011}&\multicolumn{2}{c}{2012}\\
Trigger &TOS&\multicolumn{1}{c|}{noTOS}&TOS&\multicolumn{1}{c|}{noTOS}&TOS&\multicolumn{1}{c|}{noTOS}&TOS&\multicolumn{1}{c}{noTOS}\\
\hline 
Bias data-simulation & 0.01 & 0.03 & 0.02 & 0.01 & 0.04 & 0.03 & 0.02 & 0.02 \\
\hline
Fit method & 0.00 & 0.00 & 0.00 & 0.00 & 0.00 & 0.00 & 0.00 & 0.00 \\
Kinematic acceptance & 0.03 & 0.04 & 0.02 & 0.02 & 0.06 & 0.06 & 0.06 & 0.06 \\
Resolution & 0.02 & 0.02 & 0.02 & 0.02 & 0.00 & 0.00 & 0.00 & 0.00  \\
\pwave mass model & 0.02 & 0.02 & 0.02 & 0.02 & 0.05 & 0.04 & 0.05 & 0.04 \\
\swave mass model & 0.03 & 0.03 & 0.03 & 0.03 & 0.17 & 0.17 & 0.16 & 0.17 \\
Differences data-simulation & 0.01 & 0.01 & 0.01 & 0.01 & 0.01 & 0.01  & 0.01  & 0.01 \\
Background subtraction & 0.03 & 0.03  & 0.03 & 0.03 & 0.02 & 0.01  & 0.02  & 0.01 \\
Peaking backgrounds & 0.03 & 0.04 & 0.03 & 0.04 & 0.01 & 0.01 & 0.01 & 0.01 \\
Time acceptance & $-$ & $-$ & $-$ & $-$ & 0.08 & 0.07 & 0.08 & 0.07 \\
\cdashline{1-9} 
Total systematic unc. & 0.06 & 0.08 & 0.06 & 0.07 & 0.19 & 0.19 & 0.17 & 0.18 \\
\end{tabular}
}
\end{center}
\end{table}

The inputs employed for measuring the relative branching fraction are summarised in~\tabref{tab:branching-ratio}. The factor $\kappa$ is different for the two decay modes because of two main reasons: firstly, the discrepancy between the polarisation assumed in simulation and its measurement is larger for the \BsKstKst than for the \BdKstKst decay. Secondly, the different \swave fraction of the decays. Also, the efficiency ratio of the two modes deviating from one is explained upon the different polarisation of the simulation samples. The \lhcb detector is less efficient for values of $\cos \theta_1$ ($\cos \theta_2$) close to unity because of slow pions emitted in \Kstarz (\Kstarzb) decays and these are more frequent the larger is the longitudinal polarisation.

The final result of the branching-fraction ratio is obtained as the weighted mean of the per-category result obtained with~\eqref{eq:branching-ratio} for the eight categories of~\tabref{tab:branching-ratio}, and found to be
\begin{equation}
\label{eq:branching-ratio-result}
\frac{\BF (\BdKstKst)}{\BF (\BsKstKst)} =  0.0758 \pm 0.0057 \stat \pm 0.0025 \syst \pm 0.0016 \, \left( \frac{f_s}{f_d} \right).
\end{equation}
Considering that
\[
\BF (\BsKstKst) =  (1.11 \pm 0.22 \stat \pm 0.12 \syst) \times 10^{-5},
\]
from Ref.~\cite{pdg-2018}, the absolute branching fraction for the \BdKstKst mode is found to be
\[
\BF (\BdKstKst) =  (8.0 \pm 0.9 \stat \pm 0.4 \syst) \times 10^{-7}.
\]
It is worth noticing that, since the \BsKstKst branching fraction was determined with the \decay{\Bd}{\Kstarz \phi} decay as a reference~\cite{LHCb-PAPER-2014-068}, the uncertainty on $f_s/f_d$, which appears in the ratio of~\eqref{eq:branching-ratio-result}, does not contribute to the absolute branching fraction measurement.

\begin{table}[t]
\caption{Parameters used to determine $\BF (\BdKstKst)/\BF (\BsKstKst)$. When two uncertainties are quoted, the first is statistical and the second systematic. The value of $f_s/f_d$ is taken from Ref.~\cite{fsfd}.}
\label{tab:branching-ratio}
\begin{center}
\resizebox{\textwidth}{!}{
\begin{tabular}{ l | c| c| c| c}
Parameter &  2011 TOS \MagUp & 2011 TOS \MagDown & 2011 noTOS \MagUp & 2011 noTOS \MagDown  \\
\hline
$N_{\Bd}$ &  $ \phantom{1}21.8 \pm \phantom{1}4.8 \pm 1.2$  & $ \phantom{1}33.7 \pm \phantom{1}5.5 \pm 1.4$ & $\phantom{1}10.8 \pm \phantom{1}3.6 \pm 0.9$ & $\phantom{1}33.5 \pm \phantom{1}5.4 \pm 1.4$ \\
$N_{\Bs}$ &  $145.0 \pm 10.9 \pm 3.3$ & $177.3 \pm 11.6 \pm 3.5$ & $131.9 \pm 10.5 \pm 3.2$ & $162.5 \pm 11.3 \pm 3.4$ \\
\hline
$\varepsilon_{\Bs}/\varepsilon_{\Bd}$ & $1.127 \pm 0.018 \pm 0.022$ & $1.074 \pm 0.017 \pm 0.030$ & $1.102 \pm 0.029 \pm 0.029$ & $1.144 \pm 0.030 \pm 0.026$ \\
\hline
$\kappa_{\Bd}$ &  \multicolumn{2}{c|}{$1.88 \pm 0.17 \pm 0.06$} & \multicolumn{2}{c}{$2.11 \pm 0.21 \pm 0.08$} \\
$\kappa_{\Bs}$ &  \multicolumn{2}{c|}{$3.25 \pm 0.16 \pm 0.19$} & \multicolumn{2}{c}{$3.27 \pm 0.16 \pm 0.19$} \\
\end{tabular}
}
\resizebox{\textwidth}{!}{
\begin{tabular}{ l | c| c| c| c}
\hline
Parameter &  2012 TOS \MagUp & 2012 TOS \MagDown & 2012 noTOS \MagUp & 2012 noTOS \MagDown  \\
\hline
$N_{\Bd}$ &  $ 73.0 \pm 8.7 \pm 2.3$  & $ 58.7 \pm 8.1 \pm 2.1$ & $64.1 \pm 8.4 \pm 2.2$ & $53.7 \pm 7.9 \pm 2.1$ \\
$N_{\Bs}$ &  $311\phantom{.} \pm 16\phantom{.} \pm 5\phantom{.0}$ & $344\phantom{.} \pm 17\phantom{.} \pm 5\phantom{.0}$ & $346\phantom{.} \pm 17\phantom{.} \pm 5\phantom{.0}$ & $336\phantom{.} \pm 17\phantom{.} \pm 5\phantom{.0}$ \\
\hline
$\varepsilon_{\Bs}/\varepsilon_{\Bd}$ & $1.102 \pm 0.014 \pm 0.053$ & $1.100 \pm 0.014 \pm 0.048$ & $1.180 \pm 0.022 \pm 0.065$ & $1.108 \pm 0.021 \pm 0.060$ \\
\hline
$\kappa_{\Bd}$ &  \multicolumn{2}{c|}{$1.92 \pm 0.18 \pm 0.06$} & \multicolumn{2}{c}{$2.07 \pm 0.21 \pm 0.07$} \\
$\kappa_{\Bs}$ &  \multicolumn{2}{c|}{$3.27 \pm 0.16 \pm 0.17$} & \multicolumn{2}{c}{$3.14 \pm 0.15 \pm 0.18$} \\
\hline
$f_s/f_d$ &  \multicolumn{4}{c}{$ 0.259 \pm 0.015 $} \\
\end{tabular}
}
\end{center}
\end{table}

\section{Summary and final considerations}
\label{sec:conclusion}

The first study of \BdKpiKpi decays is performed with a data set recorded by the \lhcb detector, corresponding to an integrated luminosity of 3.0\invfb at centre-of-mass energies of 7 and 8\tev. The \BdKstKst mode is observed with $10.8$ standard deviations. An untagged and time-integrated amplitude analysis is performed, taking into account the three helicity angles and the \Kpia and \Kpib invariant masses in a 150\mevcc window around the \Kstarz and \Kstarzb masses. Six contributions are included in the fit: three correspond to the \BdKstKst$\,$\pwave, and three to the \swave, along with their interferences. A large longitudinal polarisation of the \BdKstKst decay, ${f_L = 0.724 \pm 0.051 \stat \pm 0.016 \syst}$, is measured. The \swave fraction is found to be ${0.408 \pm 0.050 \stat \pm 0.023 \syst}$.

A parallel study of the \BsKpiKpi mode within $150\mevcc$ of the \Kstarz mass is performed, superseding a previous \lhcb analysis~\cite{LHCb-PAPER-2014-068}. A small longitudinal polarisation, ${f_L = 0.240 \pm 0.031 \stat \pm 0.025 \syst}$ and a large \swave contribution of ${0.694 \pm 0.016 \stat \pm 0.012 \syst}$ are measured for the \BsKstKst decay, confirming the previous \lhcb results of the time-dependent analysis of the same data~\cite{LHCb-PAPER-2017-048}.

The ratio of branching fractions
\[
{\frac{\BF (\BdKstKst)}{\BF (\BsKstKst)} = 0.0758 \pm 0.0057 \stat \pm 0.0025 \syst \pm 0.0016 \, \left(\frac{f_s}{f_d} \right)},
\]
is determined. With this ratio the \BdKstKst branching fraction is found to be
\[
\BF (\BdKstKst) =  (8.0 \pm 0.9 \stat \pm 0.4 \syst) \times 10^{-7}.
\]
This value is smaller than the measurement from the \babar collaboration~\cite{babar}, due to the \swave contribution. The measurement is compatible with the QCDF prediction of Ref.~\cite{beneke}: ${(6^{\,+1+5}_{\,-1-3})\times10^{-7}}$.

Using the \Bs-meson averages~\cite{HFLAV} for $y \equiv \Delta \Gamma/(2\Gamma)=0.064 \pm 0.005$ and the mixing phase, defined in~\eqref{eq:mixingphase}, $\phi_s=-0.021 \pm 0.031$, the ratio
\begin{equation}
\label{eq:long-BR-expression}
R_{sd} = \frac{ \BF (\BsKstKst) f_L(\BsKstKst) }{\BF(\BdKstKst)f_L(\BdKstKst)}
\frac{1-y^2}{1+y \, \cos \phi_s},
\end{equation}
is found to be 
\[
{{R_{sd} = 3.48 \pm 0.32 \stat \pm 0.19 \syst \pm 0.08 \, (f_d/f_s) \, \pm 0.02 \, (y,\phi_s)}=3.48 \pm 0.38}.
\]
This result is inconsistent with the prediction of $R_{sd} = 16.4 \pm 5.2$~\cite{DescotesGenon:2011pb}. Within models such as QCDF or the soft-collinear effective theory, based on the heavy-quark limit the predictions, longitudinal observables, such as the one in~\eqref{eq:long-BR-expression}, have reduced theoretical uncertainties as compared to parallel and perpendicular ones. The heavy-quark limit also implies the polarisation hierarchy $f_L \gg f_{\parallel,\perp}$. The measured value for $R_{sd}$ and the $f_L$ result of the \BsKstKst decay put in question this hierarchy. The picture is even more intriguing since, contrary to its U-spin partner, the \BdKstKst decay is confirmed to be strongly polarised.

\section*{Acknowledgements}
%
% These Acknowledgements valid from 3-May-2019
%
\noindent We express our gratitude to our colleagues in the CERN
accelerator departments for the excellent performance of the LHC. We
thank the technical and administrative staff at the LHCb
institutes.
We acknowledge support from CERN and from the national agencies:
CAPES, CNPq, FAPERJ and FINEP (Brazil); 
MOST and NSFC (China); 
CNRS/IN2P3 (France); 
BMBF, DFG and MPG (Germany); 
INFN (Italy); 
NWO (Netherlands); 
MNiSW and NCN (Poland); 
MEN/IFA (Romania); 
MSHE (Russia); 
MinECo (Spain); 
SNSF and SER (Switzerland); 
NASU (Ukraine); 
STFC (United Kingdom); 
DOE NP and NSF (USA).
We acknowledge the computing resources that are provided by CERN, IN2P3
(France), KIT and DESY (Germany), INFN (Italy), SURF (Netherlands),
PIC (Spain), GridPP (United Kingdom), RRCKI and Yandex
LLC (Russia), CSCS (Switzerland), IFIN-HH (Romania), CBPF (Brazil),
PL-GRID (Poland) and OSC (USA).
We are indebted to the communities behind the multiple open-source
software packages on which we depend.
Individual groups or members have received support from
AvH Foundation (Germany);
EPLANET, Marie Sk\l{}odowska-Curie Actions and ERC (European Union);
ANR, Labex P2IO and OCEVU, and R\'{e}gion Auvergne-Rh\^{o}ne-Alpes (France);
Key Research Program of Frontier Sciences of CAS, CAS PIFI, and the Thousand Talents Program (China);
RFBR, RSF and Yandex LLC (Russia);
GVA, XuntaGal and GENCAT (Spain);
the Royal Society
and the Leverhulme Trust (United Kingdom).

%\newpage
%\input{appendix_table}

\addcontentsline{toc}{section}{References}
\setboolean{inbibliography}{true}
\bibliographystyle{LHCb}
%\bibliography{main,LHCb-PAPER,LHCb-CONF,LHCb-DP,LHCb-TDR}
\bibliography{main,LHCb-PAPER,LHCb-CONF,LHCb-TDR,LHCb-DP}

\ifx\mcitethebibliography\mciteundefinedmacro
\PackageError{LHCb.bst}{mciteplus.sty has not been loaded}
{This bibstyle requires the use of the mciteplus package.}\fi
\providecommand{\href}[2]{#2}
\begin{mcitethebibliography}{10}
\mciteSetBstSublistMode{n}
\mciteSetBstMaxWidthForm{subitem}{\alph{mcitesubitemcount})}
\mciteSetBstSublistLabelBeginEnd{\mcitemaxwidthsubitemform\space}
{\relax}{\relax}

\bibitem{babar}
BaBar collaboration, B.~Aubert {\em et~al.},
  \ifthenelse{\boolean{articletitles}}{\emph{{Observation of \BdKstKst and
  search for $B^0 \rightarrow {K}^{*0}{K}^{*0}$}},
  }{}\href{https://doi.org/10.1103/PhysRevLett.100.081801}{Phys.\ Rev.\ Lett.\
  \textbf{100} (2008) 081801},
  \href{http://arxiv.org/abs/0708.2248}{{\normalfont\ttfamily
  arXiv:0708.2248}}\relax
\mciteBstWouldAddEndPuncttrue
\mciteSetBstMidEndSepPunct{\mcitedefaultmidpunct}
{\mcitedefaultendpunct}{\mcitedefaultseppunct}\relax
\EndOfBibitem
\bibitem{Chiang:2010ga}
Belle collaboration, C.-C. Chiang {\em et~al.},
  \ifthenelse{\boolean{articletitles}}{\emph{{Search for \BdKstKst, $B^0 \to
  K^{*0} K^{*0}$ and $B^0 \to K^+\pi^- K^{\mp}\pi^{\pm}$ decays}},
  }{}\href{https://doi.org/10.1103/PhysRevD.81.071101}{Phys.\ Rev.\
  \textbf{D81} (2010) 071101},
  \href{http://arxiv.org/abs/1001.4595}{{\normalfont\ttfamily
  arXiv:1001.4595}}\relax
\mciteBstWouldAddEndPuncttrue
\mciteSetBstMidEndSepPunct{\mcitedefaultmidpunct}
{\mcitedefaultendpunct}{\mcitedefaultseppunct}\relax
\EndOfBibitem
\bibitem{beneke}
M.~Beneke, J.~Rohrer, and D.~S. Yang,
  \ifthenelse{\boolean{articletitles}}{\emph{{Branching fractions, polarisation
  and asymmetries of $B \rightarrow VV$ decays}},
  }{}\href{https://doi.org/10.1016/j.nuclphysb.2007.03.020}{Nucl.\ Phys.\
  \textbf{B774} (2007) 64},
  \href{http://arxiv.org/abs/hep-ph/0612290}{{\normalfont\ttfamily
  arXiv:hep-ph/0612290}}\relax
\mciteBstWouldAddEndPuncttrue
\mciteSetBstMidEndSepPunct{\mcitedefaultmidpunct}
{\mcitedefaultendpunct}{\mcitedefaultseppunct}\relax
\EndOfBibitem
\bibitem{LHCb-PAPER-2011-012}
LHCb collaboration, R.~Aaij {\em et~al.},
  \ifthenelse{\boolean{articletitles}}{\emph{{First observation of the decay
  $\Bs\to \Kstarz\Kstarzb$}},
  }{}\href{https://doi.org/10.1016/j.physletb.2012.02.001}{Phys.\ Lett.\
  \textbf{B709} (2012) 50},
  \href{http://arxiv.org/abs/1111.4183}{{\normalfont\ttfamily
  arXiv:1111.4183}}\relax
\mciteBstWouldAddEndPuncttrue
\mciteSetBstMidEndSepPunct{\mcitedefaultmidpunct}
{\mcitedefaultendpunct}{\mcitedefaultseppunct}\relax
\EndOfBibitem
\bibitem{pdg-2018}
Particle Data Group, M.~Tanabashi {\em et~al.},
  \ifthenelse{\boolean{articletitles}}{\emph{{\href{http://pdg.lbl.gov/}{Review
  of particle physics}}},
  }{}\href{https://doi.org/10.1103/PhysRevD.98.030001}{Phys.\ Rev.\
  \textbf{D98} (2018) 030001}\relax
\mciteBstWouldAddEndPuncttrue
\mciteSetBstMidEndSepPunct{\mcitedefaultmidpunct}
{\mcitedefaultendpunct}{\mcitedefaultseppunct}\relax
\EndOfBibitem
\bibitem{LHCb-PAPER-2014-068}
LHCb collaboration, R.~Aaij {\em et~al.},
  \ifthenelse{\boolean{articletitles}}{\emph{{Measurement of \CP asymmetries
  and polarisation fractions in $\Bs\to \Kstarz\Kstarzb$ decays}},
  }{}\href{https://doi.org/10.1007/JHEP07(2015)166}{JHEP \textbf{07} (2015)
  166}, \href{http://arxiv.org/abs/1503.05362}{{\normalfont\ttfamily
  arXiv:1503.05362}}\relax
\mciteBstWouldAddEndPuncttrue
\mciteSetBstMidEndSepPunct{\mcitedefaultmidpunct}
{\mcitedefaultendpunct}{\mcitedefaultseppunct}\relax
\EndOfBibitem
\bibitem{LHCb-PAPER-2017-048}
LHCb collaboration, R.~Aaij {\em et~al.},
  \ifthenelse{\boolean{articletitles}}{\emph{{First measurement of the
  \CP-violating phase $\phi_s^{d\dquarkbar}$ in $\Bs\to(K^+\pi^-)(K^-\pi^+)$
  decays}}, }{}\href{https://doi.org/10.1007/JHEP03(2018)140}{JHEP \textbf{03}
  (2018) 140}, \href{http://arxiv.org/abs/1712.08683}{{\normalfont\ttfamily
  arXiv:1712.08683}}\relax
\mciteBstWouldAddEndPuncttrue
\mciteSetBstMidEndSepPunct{\mcitedefaultmidpunct}
{\mcitedefaultendpunct}{\mcitedefaultseppunct}\relax
\EndOfBibitem
\bibitem{Kagan:2004uw}
A.~L. Kagan, \ifthenelse{\boolean{articletitles}}{\emph{{Polarization in $B \to
  VV$ decays}}, }{}\href{https://doi.org/10.1016/j.physletb.2004.09.030}{Phys.\
  Lett.\  \textbf{B601} (2004) 151},
  \href{http://arxiv.org/abs/hep-ph/0405134}{{\normalfont\ttfamily
  arXiv:hep-ph/0405134}}\relax
\mciteBstWouldAddEndPuncttrue
\mciteSetBstMidEndSepPunct{\mcitedefaultmidpunct}
{\mcitedefaultendpunct}{\mcitedefaultseppunct}\relax
\EndOfBibitem
\bibitem{Kou:2018nap}
Belle II collaboration, E.~Kou {\em et~al.},
  \ifthenelse{\boolean{articletitles}}{\emph{{The Belle II Physics book}},
  }{}\href{http://arxiv.org/abs/1808.10567}{{\normalfont\ttfamily
  arXiv:1808.10567}}\relax
\mciteBstWouldAddEndPuncttrue
\mciteSetBstMidEndSepPunct{\mcitedefaultmidpunct}
{\mcitedefaultendpunct}{\mcitedefaultseppunct}\relax
\EndOfBibitem
\bibitem{ciucini}
M.~Ciuchini, M.~Pierini, and L.~Silvestrini,
  \ifthenelse{\boolean{articletitles}}{\emph{{$\Bs \to \kaon^{(*)0}
  \Kbar{}^{(*)0}$ \CP asymmetries: golden channels for new physics searches}},
  }{}\href{https://doi.org/10.1103/PhysRevLett.100.031802}{Phys.\ Rev.\ Lett.\
  \textbf{100} (2008) 031802},
  \href{http://arxiv.org/abs/hep-ph/0703137}{{\normalfont\ttfamily
  arXiv:hep-ph/0703137}}\relax
\mciteBstWouldAddEndPuncttrue
\mciteSetBstMidEndSepPunct{\mcitedefaultmidpunct}
{\mcitedefaultendpunct}{\mcitedefaultseppunct}\relax
\EndOfBibitem
\bibitem{Bediaga:2018lhg}
LHCb collaboration, I.~Bediaga {\em et~al.},
  \ifthenelse{\boolean{articletitles}}{\emph{{Physics case for an LHCb Upgrade
  II - Opportunities in flavour physics, and beyond, in the HL-LHC era}},
  }{}\href{http://arxiv.org/abs/1808.08865}{{\normalfont\ttfamily
  arXiv:1808.08865}}\relax
\mciteBstWouldAddEndPuncttrue
\mciteSetBstMidEndSepPunct{\mcitedefaultmidpunct}
{\mcitedefaultendpunct}{\mcitedefaultseppunct}\relax
\EndOfBibitem
\bibitem{matias}
S.~Descotes-Genon, J.~Matias, and J.~Virto,
  \ifthenelse{\boolean{articletitles}}{\emph{{Penguin-mediated $B_{d,s} \to VV$
  decays and the $B_s - \overline{B}_s$ mixing angle}},
  }{}\href{https://doi.org/10.1103/PhysRevD.76.074005}{Phys.\ Rev.\
  \textbf{D76} (2007) 074005}, Erratum
  \href{https://doi.org/10.1103/PhysRevD.84.039901}{ibid.\   \textbf{84} (2011)
  039901}, \href{http://arxiv.org/abs/0705.0477}{{\normalfont\ttfamily
  arXiv:0705.0477}}\relax
\mciteBstWouldAddEndPuncttrue
\mciteSetBstMidEndSepPunct{\mcitedefaultmidpunct}
{\mcitedefaultendpunct}{\mcitedefaultseppunct}\relax
\EndOfBibitem
\bibitem{DescotesGenon:2011pb}
S.~Descotes-Genon, J.~Matias, and J.~Virto,
  \ifthenelse{\boolean{articletitles}}{\emph{{Analysis of $B_{d,s}$ mixing
  angles in the presence of new physics and an update of $B_s \to \Kbar{}^{0*}
  \kaon^{0*}$}}, }{}\href{https://doi.org/10.1103/PhysRevD.85.034010}{Phys.\
  Rev.\  \textbf{D85} (2012) 034010},
  \href{http://arxiv.org/abs/1111.4882}{{\normalfont\ttfamily
  arXiv:1111.4882}}\relax
\mciteBstWouldAddEndPuncttrue
\mciteSetBstMidEndSepPunct{\mcitedefaultmidpunct}
{\mcitedefaultendpunct}{\mcitedefaultseppunct}\relax
\EndOfBibitem
\bibitem{Fleming:1964zz}
G.~N. Fleming, \ifthenelse{\boolean{articletitles}}{\emph{{Recoupling effects
  in the isobar model. 1. General formalism for three-pion Scattering}},
  }{}\href{https://doi.org/10.1103/PhysRev.135.B551}{Phys.\ Rev.\  \textbf{135}
  (1964) B551}\relax
\mciteBstWouldAddEndPuncttrue
\mciteSetBstMidEndSepPunct{\mcitedefaultmidpunct}
{\mcitedefaultendpunct}{\mcitedefaultseppunct}\relax
\EndOfBibitem
\bibitem{Morgan:1968zza}
D.~Morgan, \ifthenelse{\boolean{articletitles}}{\emph{{Phenomenological
  analysis of I=1/2 single-pion production processes in the energy range 500 to
  700 MeV}}, }{}\href{https://doi.org/10.1103/PhysRev.166.1731}{Phys.\ Rev.\
  \textbf{166} (1968) 1731}\relax
\mciteBstWouldAddEndPuncttrue
\mciteSetBstMidEndSepPunct{\mcitedefaultmidpunct}
{\mcitedefaultendpunct}{\mcitedefaultseppunct}\relax
\EndOfBibitem
\bibitem{Herndon:1973yn}
D.~Herndon, P.~Soding, and R.~J. Cashmore,
  \ifthenelse{\boolean{articletitles}}{\emph{{Generalized isobar model
  formalism}}, }{}\href{https://doi.org/10.1103/PhysRevD.11.3165}{Phys.\ Rev.\
  \textbf{D11} (1975) 3165}\relax
\mciteBstWouldAddEndPuncttrue
\mciteSetBstMidEndSepPunct{\mcitedefaultmidpunct}
{\mcitedefaultendpunct}{\mcitedefaultseppunct}\relax
\EndOfBibitem
\bibitem{LASS}
D.~Aston {\em et~al.}, \ifthenelse{\boolean{articletitles}}{\emph{{A study of
  $K^-\pi^+$ scattering in the reaction $K^-p \rightarrow K^-\pi^+n$ at 11
  GeV/c}}, }{}\href{https://doi.org/10.1016/0550-3213(88)90028-4}{Nucl.\ Phys.\
   \textbf{B296} (1988) 493}\relax
\mciteBstWouldAddEndPuncttrue
\mciteSetBstMidEndSepPunct{\mcitedefaultmidpunct}
{\mcitedefaultendpunct}{\mcitedefaultseppunct}\relax
\EndOfBibitem
\bibitem{bdphikst_babar}
BaBar collaboration, B.~Aubert {\em et~al.},
  \ifthenelse{\boolean{articletitles}}{\emph{{Time-dependent and
  time-integrated angular analysis of $B\rightarrow{}\phi{}{K}_{S}^{0}
  {\pi{}}^{0}$ and $\phi{}{K}^{\pm{}}{\pi{}}^{\mp{}}$}},
  }{}\href{https://doi.org/10.1103/PhysRevD.78.092008}{Phys.\ Rev.\
  \textbf{D78} (2008) 092008},
  \href{http://arxiv.org/abs/0808.3586}{{\normalfont\ttfamily
  arXiv:0808.3586}}\relax
\mciteBstWouldAddEndPuncttrue
\mciteSetBstMidEndSepPunct{\mcitedefaultmidpunct}
{\mcitedefaultendpunct}{\mcitedefaultseppunct}\relax
\EndOfBibitem
\bibitem{london_new}
B.~Bhattacharya, A.~Datta, M.~Duraisamy, and D.~London,
  \ifthenelse{\boolean{articletitles}}{\emph{{Searching for new physics with
  $\bar{b} \rightarrow \bar{s}$ $\Bs \rightarrow V_1 V_2$ penguin decays}},
  }{}\href{https://doi.org/10.1103/PhysRevD.88.016007}{{Phys.\ Rev.\ }
  \textbf{{D88}} ({2013}) {016007}},
  \href{http://arxiv.org/abs/1306.1911}{{\normalfont\ttfamily
  arXiv:1306.1911}}\relax
\mciteBstWouldAddEndPuncttrue
\mciteSetBstMidEndSepPunct{\mcitedefaultmidpunct}
{\mcitedefaultendpunct}{\mcitedefaultseppunct}\relax
\EndOfBibitem
\bibitem{HFLAV}
Heavy Flavor Averaging Group, Y.~Amhis {\em et~al.},
  \ifthenelse{\boolean{articletitles}}{\emph{{Averages of $b$-hadron,
  $c$-hadron, and $\tau$-lepton properties as of summer 2016}},
  }{}\href{https://doi.org/10.1140/epjc/s10052-017-5058-4}{Eur.\ Phys.\ J.\
  \textbf{C77} (2017) 895},
  \href{http://arxiv.org/abs/1612.07233}{{\normalfont\ttfamily
  arXiv:1612.07233}}, {updated results and plots available at
  \href{https://hflav.web.cern.ch}{{\texttt{https://hflav.web.cern.ch}}}}\relax
\mciteBstWouldAddEndPuncttrue
\mciteSetBstMidEndSepPunct{\mcitedefaultmidpunct}
{\mcitedefaultendpunct}{\mcitedefaultseppunct}\relax
\EndOfBibitem
\bibitem{Alves:2008zz}
LHCb collaboration, A.~A. Alves~Jr.\ {\em et~al.},
  \ifthenelse{\boolean{articletitles}}{\emph{{The \lhcb detector at the LHC}},
  }{}\href{https://doi.org/10.1088/1748-0221/3/08/S08005}{JINST \textbf{3}
  (2008) S08005}\relax
\mciteBstWouldAddEndPuncttrue
\mciteSetBstMidEndSepPunct{\mcitedefaultmidpunct}
{\mcitedefaultendpunct}{\mcitedefaultseppunct}\relax
\EndOfBibitem
\bibitem{LHCb-DP-2014-002}
LHCb collaboration, R.~Aaij {\em et~al.},
  \ifthenelse{\boolean{articletitles}}{\emph{{LHCb detector performance}},
  }{}\href{https://doi.org/10.1142/S0217751X15300227}{Int.\ J.\ Mod.\ Phys.\
  \textbf{A30} (2015) 1530022},
  \href{http://arxiv.org/abs/1412.6352}{{\normalfont\ttfamily
  arXiv:1412.6352}}\relax
\mciteBstWouldAddEndPuncttrue
\mciteSetBstMidEndSepPunct{\mcitedefaultmidpunct}
{\mcitedefaultendpunct}{\mcitedefaultseppunct}\relax
\EndOfBibitem
\bibitem{LHCb-DP-2012-004}
R.~Aaij {\em et~al.}, \ifthenelse{\boolean{articletitles}}{\emph{{The \lhcb
  trigger and its performance in 2011}},
  }{}\href{https://doi.org/10.1088/1748-0221/8/04/P04022}{JINST \textbf{8}
  (2013) P04022}, \href{http://arxiv.org/abs/1211.3055}{{\normalfont\ttfamily
  arXiv:1211.3055}}\relax
\mciteBstWouldAddEndPuncttrue
\mciteSetBstMidEndSepPunct{\mcitedefaultmidpunct}
{\mcitedefaultendpunct}{\mcitedefaultseppunct}\relax
\EndOfBibitem
\bibitem{Sjostrand:2006za}
T.~Sj\"{o}strand, S.~Mrenna, and P.~Skands,
  \ifthenelse{\boolean{articletitles}}{\emph{{PYTHIA 6.4 physics and manual}},
  }{}\href{https://doi.org/10.1088/1126-6708/2006/05/026}{JHEP \textbf{05}
  (2006) 026}, \href{http://arxiv.org/abs/hep-ph/0603175}{{\normalfont\ttfamily
  arXiv:hep-ph/0603175}}\relax
\mciteBstWouldAddEndPuncttrue
\mciteSetBstMidEndSepPunct{\mcitedefaultmidpunct}
{\mcitedefaultendpunct}{\mcitedefaultseppunct}\relax
\EndOfBibitem
\bibitem{LHCb-PROC-2010-056}
I.~Belyaev {\em et~al.}, \ifthenelse{\boolean{articletitles}}{\emph{{Handling
  of the generation of primary events in Gauss, the LHCb simulation
  framework}}, }{}\href{https://doi.org/10.1109/NSSMIC.2010.5873949}{Nuclear
  Science Symposium Conference Record (NSS/MIC) \textbf{IEEE} (2010)
  1155}\relax
\mciteBstWouldAddEndPuncttrue
\mciteSetBstMidEndSepPunct{\mcitedefaultmidpunct}
{\mcitedefaultendpunct}{\mcitedefaultseppunct}\relax
\EndOfBibitem
\bibitem{Lange:2001uf}
D.~J. Lange, \ifthenelse{\boolean{articletitles}}{\emph{{The EvtGen particle
  decay simulation package}},
  }{}\href{https://doi.org/10.1016/S0168-9002(01)00089-4}{Nucl.\ Instrum.\
  Meth.\  \textbf{A462} (2001) 152}\relax
\mciteBstWouldAddEndPuncttrue
\mciteSetBstMidEndSepPunct{\mcitedefaultmidpunct}
{\mcitedefaultendpunct}{\mcitedefaultseppunct}\relax
\EndOfBibitem
\bibitem{Golonka:2005pn}
P.~Golonka and Z.~Was, \ifthenelse{\boolean{articletitles}}{\emph{{PHOTOS Monte
  Carlo: A precision tool for QED corrections in $Z$ and $W$ decays}},
  }{}\href{https://doi.org/10.1140/epjc/s2005-02396-4}{Eur.\ Phys.\ J.\
  \textbf{C45} (2006) 97},
  \href{http://arxiv.org/abs/hep-ph/0506026}{{\normalfont\ttfamily
  arXiv:hep-ph/0506026}}\relax
\mciteBstWouldAddEndPuncttrue
\mciteSetBstMidEndSepPunct{\mcitedefaultmidpunct}
{\mcitedefaultendpunct}{\mcitedefaultseppunct}\relax
\EndOfBibitem
\bibitem{Allison:2006ve}
Geant4 collaboration, J.~Allison {\em et~al.},
  \ifthenelse{\boolean{articletitles}}{\emph{{Geant4 developments and
  applications}}, }{}\href{https://doi.org/10.1109/TNS.2006.869826}{IEEE
  Trans.\ Nucl.\ Sci.\  \textbf{53} (2006) 270}\relax
\mciteBstWouldAddEndPuncttrue
\mciteSetBstMidEndSepPunct{\mcitedefaultmidpunct}
{\mcitedefaultendpunct}{\mcitedefaultseppunct}\relax
\EndOfBibitem
\bibitem{Agostinelli:2002hh}
Geant4 collaboration, S.~Agostinelli {\em et~al.},
  \ifthenelse{\boolean{articletitles}}{\emph{{Geant4: a simulation toolkit}},
  }{}\href{https://doi.org/10.1016/S0168-9002(03)01368-8}{Nucl.\ Instrum.\
  Meth.\  \textbf{A506} (2003) 250}\relax
\mciteBstWouldAddEndPuncttrue
\mciteSetBstMidEndSepPunct{\mcitedefaultmidpunct}
{\mcitedefaultendpunct}{\mcitedefaultseppunct}\relax
\EndOfBibitem
\bibitem{LHCb-PROC-2011-006}
M.~Clemencic {\em et~al.}, \ifthenelse{\boolean{articletitles}}{\emph{{The
  \lhcb simulation application, Gauss: Design, evolution and experience}},
  }{}\href{https://doi.org/10.1088/1742-6596/331/3/032023}{{J.\ Phys.\ Conf.\
  Ser.\ } \textbf{331} (2011) 032023}\relax
\mciteBstWouldAddEndPuncttrue
\mciteSetBstMidEndSepPunct{\mcitedefaultmidpunct}
{\mcitedefaultendpunct}{\mcitedefaultseppunct}\relax
\EndOfBibitem
\bibitem{Breiman}
L.~Breiman, J.~H. Friedman, R.~A. Olshen, and C.~J. Stone, {\em Classification
  and regression trees}, Wadsworth international group, Belmont, California,
  USA, 1984\relax
\mciteBstWouldAddEndPuncttrue
\mciteSetBstMidEndSepPunct{\mcitedefaultmidpunct}
{\mcitedefaultendpunct}{\mcitedefaultseppunct}\relax
\EndOfBibitem
\bibitem{Roe}
B.~P. Roe {\em et~al.}, \ifthenelse{\boolean{articletitles}}{\emph{{Boosted
  decision trees as an alternative to artificial neural networks for particle
  identification}}, }{}\href{https://doi.org/10.1016/j.nima.2004.12.018}{Nucl.\
  Instrum.\ Meth.\  \textbf{A543} (2005) 577},
  \href{http://arxiv.org/abs/physics/0408124}{{\normalfont\ttfamily
  arXiv:physics/0408124}}\relax
\mciteBstWouldAddEndPuncttrue
\mciteSetBstMidEndSepPunct{\mcitedefaultmidpunct}
{\mcitedefaultendpunct}{\mcitedefaultseppunct}\relax
\EndOfBibitem
\bibitem{MartinezSantos2014150}
D.~Mart\'inez~Santos and F.~Dupertuis,
  \ifthenelse{\boolean{articletitles}}{\emph{{Mass distributions marginalized
  over per-event errors}},
  }{}\href{https://doi.org/10.1016/j.nima.2014.06.081}{Nucl.\ Instrum.\ Meth.\
  \textbf{A764} (2014) 150},
  \href{http://arxiv.org/abs/1312.5000}{{\normalfont\ttfamily
  arXiv:1312.5000}}\relax
\mciteBstWouldAddEndPuncttrue
\mciteSetBstMidEndSepPunct{\mcitedefaultmidpunct}
{\mcitedefaultendpunct}{\mcitedefaultseppunct}\relax
\EndOfBibitem
\bibitem{Skwarnicki:1986xj}
T.~Skwarnicki, {\em {A study of the radiative cascade transitions between the
  Upsilon-prime and Upsilon resonances}}, PhD thesis, Institute of Nuclear
  Physics, Krakow, 1986,
  {\href{http://inspirehep.net/record/230779/}{DESY-F31-86-02}}\relax
\mciteBstWouldAddEndPuncttrue
\mciteSetBstMidEndSepPunct{\mcitedefaultmidpunct}
{\mcitedefaultendpunct}{\mcitedefaultseppunct}\relax
\EndOfBibitem
\bibitem{Albrecht:1990cs}
ARGUS collaboration, H.~Albrecht {\em et~al.},
  \ifthenelse{\boolean{articletitles}}{\emph{{Exclusive hadronic decays of $B$
  Mesons}}, }{}\href{https://doi.org/10.1007/BF01614687}{Z.\ Phys.\
  \textbf{C48} (1990) 543}\relax
\mciteBstWouldAddEndPuncttrue
\mciteSetBstMidEndSepPunct{\mcitedefaultmidpunct}
{\mcitedefaultendpunct}{\mcitedefaultseppunct}\relax
\EndOfBibitem
\bibitem{Pivk:2004ty}
M.~Pivk and F.~R. Le~Diberder,
  \ifthenelse{\boolean{articletitles}}{\emph{{sPlot: a statistical tool to
  unfold data distributions}},
  }{}\href{https://doi.org/10.1016/j.nima.2005.08.106}{Nucl.\ Instrum.\ Meth.\
  \textbf{A555} (2005) 356},
  \href{http://arxiv.org/abs/physics/0402083}{{\normalfont\ttfamily
  arXiv:physics/0402083}}\relax
\mciteBstWouldAddEndPuncttrue
\mciteSetBstMidEndSepPunct{\mcitedefaultmidpunct}
{\mcitedefaultendpunct}{\mcitedefaultseppunct}\relax
\EndOfBibitem
\bibitem{Xie:2009rka}
Y.~Xie, \ifthenelse{\boolean{articletitles}}{\emph{{sFit: a method for
  background subtraction in maximum likelihood fit}},
  }{}\href{http://arxiv.org/abs/0905.0724}{{\normalfont\ttfamily
  arXiv:0905.0724}}\relax
\mciteBstWouldAddEndPuncttrue
\mciteSetBstMidEndSepPunct{\mcitedefaultmidpunct}
{\mcitedefaultendpunct}{\mcitedefaultseppunct}\relax
\EndOfBibitem
\bibitem{TristansThesis}
T.~du~Pree, {\em {Search for a strange phase in beautiful oscillations}}, PhD
  thesis, {Vrije Universiteit Amsterdam}, 2010,
  {\href{https://cds.cern.ch/record/1299931}{CERN-THESIS-2010-124}}\relax
\mciteBstWouldAddEndPuncttrue
\mciteSetBstMidEndSepPunct{\mcitedefaultmidpunct}
{\mcitedefaultendpunct}{\mcitedefaultseppunct}\relax
\EndOfBibitem
\bibitem{LHCb-PAPER-2014-014}
LHCb collaboration, R.~Aaij {\em et~al.},
  \ifthenelse{\boolean{articletitles}}{\emph{{Observation of the resonant
  character of the $Z(4430)^-$ state}},
  }{}\href{https://doi.org/10.1103/PhysRevLett.112.222002}{Phys.\ Rev.\ Lett.\
  \textbf{112} (2014) 222002},
  \href{http://arxiv.org/abs/1404.1903}{{\normalfont\ttfamily
  arXiv:1404.1903}}\relax
\mciteBstWouldAddEndPuncttrue
\mciteSetBstMidEndSepPunct{\mcitedefaultmidpunct}
{\mcitedefaultendpunct}{\mcitedefaultseppunct}\relax
\EndOfBibitem
\bibitem{Pelaez:2016tgi}
J.~R. Pel\'aez and A.~Rodas,
  \ifthenelse{\boolean{articletitles}}{\emph{{Pion-kaon scattering amplitude
  constrained with forward dispersion relations up to 1.6 GeV}},
  }{}\href{https://doi.org/10.1103/PhysRevD.93.074025}{Phys.\ Rev.\
  \textbf{D93} (2016) 074025},
  \href{http://arxiv.org/abs/1602.08404}{{\normalfont\ttfamily
  arXiv:1602.08404}}\relax
\mciteBstWouldAddEndPuncttrue
\mciteSetBstMidEndSepPunct{\mcitedefaultmidpunct}
{\mcitedefaultendpunct}{\mcitedefaultseppunct}\relax
\EndOfBibitem
\bibitem{GarciaPardinas:2630181}
J.~Garc\'{\i}a Pardi\~{n}as, {\em {Search for flavour anomalies at LHCb:
  decay-time-dependent \CP violation in $B_s^0\to(K^+\pi^-)(K^-\pi^+)$ and
  lepton universality in $\overline{B}{}^0\to D^{(*)+}l\overline{\nu}_{l}$}},
  PhD thesis, {Universidade de Santiago de Compostela}, 2018,
  {\href{https://cds.cern.ch/record/2630181}{CERN-THESIS-2018-096}}\relax
\mciteBstWouldAddEndPuncttrue
\mciteSetBstMidEndSepPunct{\mcitedefaultmidpunct}
{\mcitedefaultendpunct}{\mcitedefaultseppunct}\relax
\EndOfBibitem
\bibitem{LHCb-DP-2012-003}
M.~Adinolfi {\em et~al.},
  \ifthenelse{\boolean{articletitles}}{\emph{{Performance of the \lhcb RICH
  detector at the LHC}},
  }{}\href{https://doi.org/10.1140/epjc/s10052-013-2431-9}{Eur.\ Phys.\ J.\
  \textbf{C73} (2013) 2431},
  \href{http://arxiv.org/abs/1211.6759}{{\normalfont\ttfamily
  arXiv:1211.6759}}\relax
\mciteBstWouldAddEndPuncttrue
\mciteSetBstMidEndSepPunct{\mcitedefaultmidpunct}
{\mcitedefaultendpunct}{\mcitedefaultseppunct}\relax
\EndOfBibitem
\bibitem{LHCb-PUB-2016-021}
L.~Anderlini {\em et~al.}, \ifthenelse{\boolean{articletitles}}{\emph{{The
  PIDCalib package}}, }{}
  \href{http://cdsweb.cern.ch/search?p=LHCb-PUB-2016-021&f=reportnumber&action_search=Search&c=LHCb+Notes}
  {LHCb-PUB-2016-021}, 2016\relax
\mciteBstWouldAddEndPuncttrue
\mciteSetBstMidEndSepPunct{\mcitedefaultmidpunct}
{\mcitedefaultendpunct}{\mcitedefaultseppunct}\relax
\EndOfBibitem
\bibitem{fsfd}
LHCb collaboration, R.~Aaij {\em et~al.},
  \ifthenelse{\boolean{articletitles}}{\emph{{Measurement of the fragmentation
  fraction ratio $f_s/f_d$ and its dependence on $B$ meson kinematics}},
  }{}\href{https://doi.org/10.1007/JHEP04(2013)001}{JHEP \textbf{04} (2013)
  001}, \href{http://arxiv.org/abs/1301.5286}{{\normalfont\ttfamily
  arXiv:1301.5286}}, $f_s/f_d$ value updated in
  \href{https://cds.cern.ch/record/1559262}{LHCb-CONF-2013-011}\relax
\mciteBstWouldAddEndPuncttrue
\mciteSetBstMidEndSepPunct{\mcitedefaultmidpunct}
{\mcitedefaultendpunct}{\mcitedefaultseppunct}\relax
\EndOfBibitem
\end{mcitethebibliography}
%\bibliography{main}

\newpage
% LHCb Collaboration author list
% Data extracted on April 24th, 2019 at 3:47pm for reference date 25-Jan-2019
\centerline
{\large\bf LHCb Collaboration}
\begin
{flushleft}
\small
R.~Aaij$^{29}$,
C.~Abell{\'a}n~Beteta$^{46}$,
B.~Adeva$^{43}$,
M.~Adinolfi$^{50}$,
C.A.~Aidala$^{77}$,
Z.~Ajaltouni$^{7}$,
S.~Akar$^{61}$,
P.~Albicocco$^{20}$,
J.~Albrecht$^{12}$,
F.~Alessio$^{44}$
% ...

{\footnotesize \it

% ...
$ ^{7}$Universit{\'e} Clermont Auvergne, CNRS/IN2P3, LPC, Clermont-Ferrand, France\\
% ...
$ ^{12}$Fakult{\"a}t Physik, Technische Universit{\"a}t Dortmund, Dortmund, Germany\\
% ...
$ ^{20}$INFN Laboratori Nazionali di Frascati, Frascati, Italy\\
% ...
$ ^{29}$Nikhef National Institute for Subatomic Physics, Amsterdam, Netherlands\\
% ...
$ ^{43}$Instituto Galego de F{\'\i}sica de Altas Enerx{\'\i}as (IGFAE), Universidade de Santiago de Compostela, Santiago de Compostela, Spain\\
$ ^{44}$European Organization for Nuclear Research (CERN), Geneva, Switzerland\\
% ...
$ ^{46}$Physik-Institut, Universit{\"a}t Z{\"u}rich, Z{\"u}rich, Switzerland\\
% ...
$ ^{50}$H.H. Wills Physics Laboratory, University of Bristol, Bristol, United Kingdom\\
% ...
$ ^{61}$University of Cincinnati, Cincinnati, OH, United States\\
% ...
$ ^{63}$Syracuse University, Syracuse, NY, United States\\
% ...
$ ^{77}$University of Michigan, Ann Arbor, United States, associated to $^{63}$\\
% ...
}
\end{flushleft}

\end{document}